\documentclass[10pt,letterpaper]{article}
\usepackage[top=0.85in,left=2.75in,footskip=0.75in]{geometry}

\usepackage{amsmath,amssymb}

\usepackage{changepage}

\usepackage{textcomp,marvosym}

\usepackage{cite}

\usepackage{nameref,hyperref}

\usepackage[right]{lineno}

\usepackage[nopatch=eqnum]{microtype}
\DisableLigatures[f]{encoding = *, family = * }

\usepackage[table]{xcolor}

\usepackage{array}

\newcolumntype{+}{!{\vrule width 2pt}}

\newlength\savedwidth

\raggedright
\usepackage[aboveskip=1pt,labelfont=bf,labelsep=period,justification=raggedright,singlelinecheck=off]{caption}

\makeatletter
\renewcommand{\@biblabel}[1]{\quad#1.}
\makeatother

\usepackage{lastpage,fancyhdr,graphicx}
\usepackage{epstopdf}
\fancyheadoffset[L]{2.25in}
\fancyfootoffset[L]{2.25in}
\usepackage{siunitx}
\usepackage{graphicx}
\usepackage{subcaption}
\usepackage{bm}
\newcommand{\param}{\bm{\theta}}

\newcommand{\data}{\bm{Y}}
\newcommand{\numpcecoeff}{\mathcal{K}}
\newcommand{\pcecoeff}{z}
\newcommand{\pcepoly}{\phi}
\newcommand{\pcemono}{\psi}
\newcommand{\expect}{\textbf{E}}
\newcommand{\var}{\textbf{Var}}
\newcommand{\pcecoeffmat}{\bm{Z}}
\newcommand{\pcepolymat}{\bm{\Phi}}
\newcommand{\numparam}{P}
\newcommand{\pceorder}{m}

\begin{document}
\vspace*{0.2in}

\begin{flushleft}
{\Large
\textbf\newline{Assessing parameter identifiability of a hemodynamics PDE model using spectral surrogates and dimension reduction} 
}
\newline
\\
Mitchel J. Colebank\textsuperscript{1,2*},
\\
\bigskip
\textbf{1} Department of Mathematics, University of South Carolina, Columbia, SC, 29208, USA
\\
\textbf{2} Department of Biomedical Engineering, University of South Carolina, Columbia, SC, 29208, USA
\\
\bigskip


* mjcolebank@sc.edu

\end{flushleft}
\section*{Abstract}
Computational inverse problems for biomedical simulators suffer from limited data and relatively high parameter dimensionality. This often requires sensitivity analysis, where parameters of the model are ranked based on their influence on the specific quantities of interest. This is especially important for simulators used to build medical digital twins, as the amount of data is typically limited. For expensive models, such as blood flow models, emulation is employed to expedite the simulation time. Parameter ranking and fixing using sensitivity analysis are often heuristic, though, and vary with the specific application or simulator used. The present study provides an innovative solution to this problem by leveraging polynomial chaos expansions (PCEs) for both multioutput global sensitivity analysis and formal parameter identifiability. For the former, we use dimension reduction to efficiently quantify time-series sensitivity of a one-dimensional pulmonary hemodynamics model. We consider both Windkessel and structured tree boundary conditions. We then use PCEs to construct profile-likelihood confidence intervals to formally assess parameter identifiability, and show how changes in experimental design improve identifiability. Our work presents a novel approach to determining parameter identifiability and leverages a common emulation strategy for enabling profile-likelihood analysis in problems governed by partial differential equations.

\section*{Author summary}
The calibration of biophysical models is often ill-posed, with the parameter dimensionality typically larger than available data for parameter inference. In addition, these models often employ parameters that are clinically or experimentally interpretable, hence a unique set of estimated parameters are necessary for interpreting biophysical processes. Sensitivity analysis is a necessary tool for reducing the parameter dimensionality by ``fixing'' noninfluential parameters, yet choosing the cutoff for parameter fixing is problem dependent. Identifiability methods like profile-likelihood are computationally expensive, and have traditionally been reserved for relatively fast simulators. We show that emulation, using polynomial chaos, provides a framework for a two-in-one analysis of model sensitivity and parameter identifiability. Using a pulmonary hemodynamics simulator, we show how this framework allows for a more formal analysis of the model and its parameters. Our approach allows us to examine how different measurement modalities affect the ability to infer biophysical parameters, and is a step forward in developing data-specific models for understanding cardiovascular disease.


\section*{Introduction}
Computational modeling and simulation is a cornerstone of the digital twin pipeline \cite{Kimpton2025,Colebank2024G}, which is now considered the next scientific frontier in personalized medicine. As documented extensively in the report from the National Academies of Sciences, Engineering, and Medicine \cite{NatAcadDT}, the development of digital twins requires proper uncertainty quantification (UQ) for informed decision making. Given that the mathematical simulators, which are often expensive, need to be made data-specific by solving an inverse problem, there is an inherent need for methods that identify which parameters should be prioritized for inference. Biophysical problems typically prescribe meaning to the parameters themselves, and thus inferred parameters may be viewed as an additional feature for personalized medicine \cite{Kimpton2025,Colebank2024G}. 

Global sensitivity analysis includes moment and moment-independent methods \cite{Eck2016,Smith2024,Borgonovo2016} and seek to explain how uncertainties in input parameters correspond to uncertainties in model outputs. Moment based methods are more commonly employed, and utilize the concepts of expectation (i.e., the average) and variance to quantify how parameters contribute to the variance of the output \cite{Eck2016}. Moment independent methods assess how the entire probability distribution function (PDF) of the output(s) are shifted or changed in response to changes in the model inputs \cite{Borgonovo2016}. The most popular moment-based sensitivity measure is variance-based Sobol' indices \cite{Smith2024}, which attribute proportions of the variance in output space to parameters and their interactions. These methods have been extended to time-dependent \cite{Alexanderian2020} and multivariate \cite{Nagel2020} outputs. In the case of expensive simulators, e.g. blood flow models dictated by partial differential equations (PDEs), surrogate models (or \textit{emulators}) are used \cite{Eck2016,Eck2017,Huberts2014,Colebank2024}. Polynomial chaos expansions (PCEs) are a special type of spectral polynomial surrogate that are designed to be orthogonal with respect to a prior probability measure \cite{Eck2016,Borgonovo2016}, making them efficient for calculating Sobol' indices. Parameters with little to no contribution to the variance of the system are deemed ``non-influential'', and can be fixed at \textit{apriori} values while influential parameters are prioritized for inference. Non-influential parameters are considered ``practically non-identifiable'' in that they have little to no impact on the model output and will be difficult to infer from data given an experimental design \cite{Smith2024}. Yet, a robust cutoff for which parameters are non-influential are unavailable and typically heuristic. Thus, new methods for determining when parameters are both influential and identifiable are needed. The profile-likelihood is a powerful identifiability method, which assesses which parameters are identifiable by constructing confidence intervals for each parameter one at a time \cite{Colebank2024G,Wieland2021,Colebank2022}. If confidence intervals are bounded, then the parameter is considered identifiable, whereas parameters with only one or no finite confidence bounds are deemed (practically) non-identifiable. This approach has been used in multiple applications \cite{Colebank2022,Wieland2021}, yet only several studies have assessed identifiability through this method with PDE simulators \cite{Renardy2022structural,Murphy2024implementing,Ciocanel2024parameter}. This is largely due to the computational cost of constructing profile-likelihood confidence intervals, which require solving multiple optimization problems for the profiled parameter. Hence computational cost for PDE simulators is a bottleneck for using such an approach.

We bridge this knowledge gap by using spectral PCE surrogates to quantify model sensitivity using Sobol' indices, and then test for parameter identifiability using profile-likelihood confidence intervals. We consider a one-dimensional (1D) pulse-wave propagation model of the pulmonary circulation with two different classes of boundary conditions: Windkessel models and structured tree models of the distal vasculature \cite{Paun2025,Olufsen2000}. We begin by presenting a PCE surrogate methodology that combines dimension reduction, via principal component analysis (PCA), to emulate the time-dependent PDE outputs. We then illustrate how Sobol' indices for both PCA and time-dependent outputs can be calculated using the PCE coefficients \cite{Nagel2020}. The PCEs, built on the PCA representation of the output, are then used to calculate profile-likelihood confidence intervals for each parameter at a significantly lower computational cost. We consider three different experimental designs for our models, and illustrate how sensitivity and identifiability analyses can be used to probe model parameters and determine which parameters should be fixed given an experimental design. Our results directly apply to the field of computational hemodynamics modeling, but also provide a framework for future analyses of PDE models by combining PCEs, global sensitivity analysis, and identifiability analysis.

\section*{Methods}

\subsection*{Fluid Dynamics Model}\label{sec:supp_fluids}
The computational hemodynamics model is a simplified version of the three-dimensional Navier-Stokes equations in cylindrical coordinates. The model simulates pulse-wave propagation throughout a network of blood vessels under the assumption that blood flow is laminar, Newtonian, incompressible, and axially dominant \cite{Colebank2019,Colebank2024,Paun2020}. In addition, we assume that blood vessels are cylindrical and impermeable. The system constitutes a set of nonlinear, hyperbolic, PDEs, given by the mass conservation and momentum balance equations
\begin{align}\label{eq:mass}
    \displaystyle\frac{\partial A}{\partial t} + \frac{\partial q}{\partial x} &= 0, \\
    \label{eq:momentum}       
    \displaystyle\frac{\partial q}{\partial t} + \left(\frac{\gamma+2}{\gamma+1}\right)\frac{\partial}{\partial x}\left(\frac{q^2}{A}\right) + \frac{A}{\rho}\frac{\partial p}{\partial x} &= -\frac{2\pi \mu (\gamma+2)}{\rho} \left(\frac{q}{A}\right),
\end{align}
respectively. The above equation describes the interactions between blood pressure $p(x,t)$, (mmHg), cross-sectional area $A(x,t)$, (cm$^2$) and volumetric flow rate $q(x,t)$, (mL/s). The inertial and viscous shear stress terms in \eqref{eq:momentum} are derived from the assumed velocity ($u=q/A$, cm/s) profile
\begin{equation}
u(r,x,t) = U(x,t)\frac{\gamma+2}{\gamma}\left(1-\left(\frac{r}{R(x,t)}\right)^{\gamma}\right),
\end{equation}
where $U(x,t)$ (cm/s) is the average axial velocity and $\gamma$ is the power-law exponent. To obtain a relatively flat velocity profile as seen in-vivo, we set $\gamma=9$ \cite{VandeVosse2011}. Finally, we relate pressure and area using the linear stress-strain relationship
\begin{equation}\label{eq:state}
    p(x,t) = \frac{4}{3}\frac{E h}{r_0}\left(\sqrt{\frac{A}{A_0}}-1\right),  \ \ \  \frac{Eh}{r_0}= k_1 e^{-k_2 r_0} + k_3
\end{equation}
where $A_0 = \pi r_0^2$ (cm$^2$) is the reference area, $E$ (mmHg) is the Young's modulus, and $h$ (cm) is the wall thickness. The coefficient $Eh/r_0$ is assumed to follow the above exponential relationship where $k_1$ (mmHg), $k_2$(cm$^{-1}$), and $k_3$ (mmHg) are parameters \cite{Olufsen2000}. 

\begin{figure}
    \includegraphics[width=\linewidth]{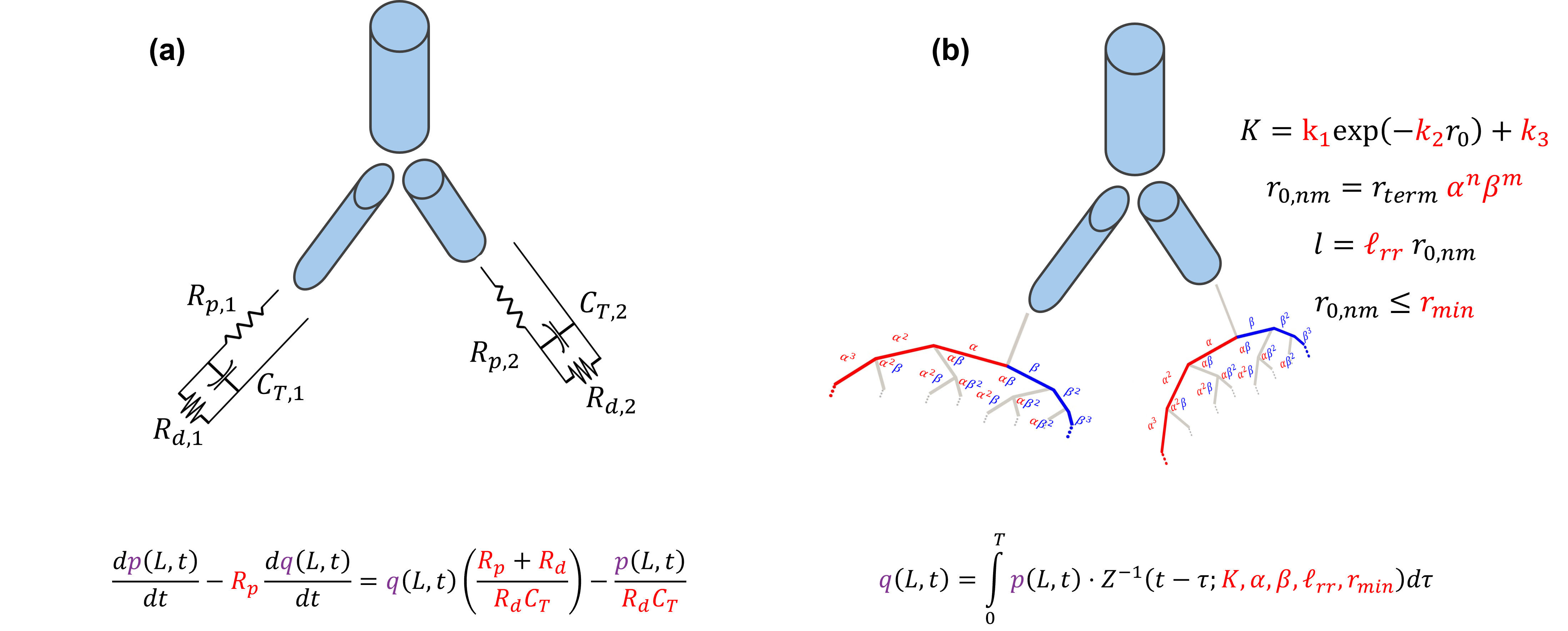}
    \caption{Schematic of the mathematical models, including three large pulmonary vessels attached to either (a) Windkessel boundary conditions or (b) structured tree boundary conditions. The variables in red denote the parameters to be inferred, as well as their mathematical representations in the boundary conditions. Variables in purple are state variables of the model.}
    \label{fig:Schematic}
\end{figure}

We consider a relatively small network in this study, including the main, left, and right pulmonary arteries (MPA, LPA, and RPA, respectively), using the dimensions presented in Qureshi et al. \cite{Qureshi2014}. We use a pulmonary blood flow time-series boundary condition at the inlet of the MPA that is representative of MPA flow magnitude and shape \cite{Yang2019}. We enforce continuity of total pressure and conservation of flow at each vascular junction. At the distal end of the large vessels, we enforce one of two boundary conditions, as discussed below and shown in Figure \ref{fig:Schematic}. The model is solved using a two-step Lax-Wendroff finite difference scheme \cite{Olufsen2000}.

\subsection*{Windkessel Boundary Conditions}\label{sec:supp_WK}
Windkessel models are a three-element electrical circuit \cite{Westerhof2009} which represent the proximal and distal resistances to blood flow as well as a total compliance, represented by $R_p$ (mmHg s/ ml), $R_d$ (mmHg s/ ml), and  $C_T$ (ml/mmHg) respectively. This circuit model is mathematically represented by the first order differential equation
\begin{equation}
\frac{dp}{dt} = R_p \left(\frac{dq}{dt}\right)+q\left(\frac{R_p+R_d}{R_pR_d}\right) - \frac{p}{R_d C_T}. \label{eq:Windkessel}
\end{equation}
The Windkessel models are attached to the two terminal vessels in the network, each with their own unique $R_p,R_d,$ and $C_T$ value. Hence the full parameter set for the Windkessel model is
\begin{equation}\label{eq:WK_param}
    \param_{WK} = \left\{k_1,k_2,k_3,R_{p,1},R_{p,2},R_{d,1},R_{p,2},C_{T,1},C_{T,2}\right\}
\end{equation}
where the subscripts $1$ and $2$ denote the left and right pulmonary artery boundary conditions, respectively.

\subsection*{Structured Tree Boundary Conditions} \label{sec:supp_ST}
The structured tree model \cite{Olufsen2000} is an alternative boundary condition that calculates an asymmetric, bifurcating, synthetic vascular tree at the end of the two terminal vessels. The structured tree includes a large radii scaling factor, $\alpha$, a small radii scaling factor, $\beta$, a length-to-radius ratio, $\ell_{rr}$, and a minimum radius for the structured tree, $r_{min}$ (cm). Additional details can be found in \cite{Chambers2020,Olufsen2000}. Whereas the Windkessel model has a unique set of parameters for each terminal branch, we assume a common set of parameters for both the LPA and RPA when using this boundary condition. The coupling between the PDE system and the structured tree boundary condition is defined by the convolution integral
\begin{equation}
    q(L,t) = \int_{0}^{t} p(L,\tau)\cdot Z^{-1}\left(t-\tau;\alpha,\beta,\ell_{rr},r_{min}\right) d\tau
\end{equation}
where $x=L$ is the final spatial point in the vessel and $Z(t;\alpha,\beta,\ell_{rr},r_{min})$ is the total impedance of the structured tree, which is calculated under the assumptions of viscous dominant, periodic flow and the parameterization of the structured tree geometry itself \cite{Olufsen2000}. Thus, the full parameter set for the structured tree model is
\begin{equation}\label{eq:ST_param}
    \param_{ST} = \left\{k_1,k_2,k_3,\alpha,\beta,\ell_{rr},r_{min}\right\}.
\end{equation}

\subsection*{Spectral Surrogate}\label{sec:PCEs}
Polynomial chaos expansions (PCEs) are a spectral surrogate modeling technique that use standard polynomial types that are orthogonal with respect to a prior probability distribution measure \cite{Eck2016}. Let $\data = \mathcal{M}(\bm{\theta})$ denote the output quantity of interest given by the model $\mathcal{M}$ for some $P$ dimensional input parameter vector $\bm{\theta}\in\mathbb{R}^P$. The PCE approximates the function using the finite summation
\begin{equation}
    \mathcal{M}(\theta) \approx \mathcal{M}^{\text{PCE}}(\param) =  \sum_{j=0}^{\numpcecoeff-1} \pcecoeff_j \pcepoly_j(\param), \ \ \ \pcepoly_j(\param) = \prod_{i=1}^{\numparam}\pcemono_{ij}(\theta_i),
\end{equation}
where $\pcecoeff_j$ are the polynomial coefficients corresponding to the multivariate polynomials $\pcepoly_j(\param)$. These multivariate polynomials are formed by the product of univariate polynomials $\pcemono_{ij}(\theta_i)$ for polynomial basis function $j$. The polynomial type is associated with the prior distribution for each polynomial; e.g., uniform distributions (used here) correspond to Legendre polynomials \cite{Eck2016}. The total number of polynomial basis functions is denoted by $\numpcecoeff = {(\numparam+\pceorder) \choose \pceorder}$, which scales with $\numparam$ parameters and the polynomial order of $\pceorder$.

As mentioned above, the PCE polynomials are chosen to be orthogonal with respect to the prior probability measure, giving
\begin{equation}
    \expect\left[\pcepoly_j(\param),\pcepoly_{j'}(\param)\right] = \int_{\Gamma} \pcepoly_j(\bm{\theta}) \pcepoly_{j'}(\bm{\theta})\rho_{\theta}d\theta = \gamma_j\cdot \delta_{j,j'},
\end{equation}
where $\rho_{\param}$ is the prior parameter density. The normalization factor $\gamma_j = \expect\left[\pcepoly_j^2\right]$ is specific to the polynomial type. We assume uniform priors on all our parameters (after mapping them to the interval $[-1,1]$ \cite{Huberts2014}), and subsequently use Legendre polynomials, with $\gamma_j = 2/(2j+1)$.

We determine the polynomial coefficients, $\pcecoeffmat$ using non-intrusive regression \cite{Eck2016}. Given the simulator outputs $\data$, the coefficients of the PCE are calculated using the ordinary least squares solutions
\begin{equation}
    \pcecoeffmat = \left(\pcepolymat^\top \pcepolymat\right)^{-1}\pcepolymat^{\top}\data,
\end{equation}
where $\pcepolymat$ is the matrix of polynomials evaluated at each parameter value. The size of both the polynomial matrix $\pcepolymat$ and the coefficient matrix $\pcecoeffmat$ is dictated by the size of the solution space in $\data$. We use the UQlab software in MATLAB for PCE construction and evaluation \cite{marelli2014uqlab}.

\subsection*{Dimensionality reduction}
The blood flow PDE model includes three dynamic states: pressure, flow, and area. The states are spatiotemporal signals in the three blood vessels. We consider model outputs at the midpoint of each vessel, thus leaving temporal signals. Rather than calibrate a PCE to these true PDE signals, we emulate on a reduced representation of the multivariate simulator output using PCA \cite{Higdon2008,Paun2025}. Similar to proper orthogonal decomposition (POD) and following from the general decomposition via the Karhunen–Loève expansion \cite{Alexanderian2020}, PCA constructs a reduced subspace of the original data using a singular value decomposition of the data covariance matrix. As done previously for this model \cite{Paun2025}, the simulator output can be decomposed via
\begin{equation}\label{eq:PCA}
Y(\bm{\theta}, \textbf{t}) \approx \boldsymbol{\mu}(\textbf{t}) + \sum_{q=1}^M c_q(\bm{\theta}, \textbf{t}) \eta_q(t).
\end{equation}
The above decomposition accounts for the time-dependent average, $\expect\left[Y(\bm{\theta}, \textbf{t})\right] = \boldsymbol{\mu}(\textbf{t})$, of the model response, as well as the principal component scores, $c_q(\bm{\theta}$ and orthonormal basis vectors $\eta_q$. The orthonormal basis vectors provide a rotated coordinate frame for the original model response, and can be used to construct a lower dimensional representation for the output domain, i.e. $M<N_t$, where $N_t$ is the number of time-points in the original signals and $M$ is the selected number of principal components. 
For $M$ principal components, the PCE requires  $\pcepolymat\in \mathbb{R}^{\numpcecoeff\times M}$ polynomials. The assumption of independent outputs in the PCA representation encourages the use of independent PCEs for the coefficients. The PCE polynomial coefficients are computed by regressing on the PCA scores
\begin{equation} \label{eq:pce_PCA_first}
    c_q(\bm{\theta}) = \sum_{k=0}^{\numpcecoeff-1} \pcecoeff_{qk} \pcepoly_k(\bm{\theta}), q=1,\dots,M
\end{equation}
for each principal component. The PCE predictions map the parameter vector using the polynomial bases and coefficients defined in eq~\eqref{eq:pce_PCA_first} to the PCA scores, which can then be transformed into the time-domain using eq~\eqref{eq:PCA}. Finally, we transform the PCE based surrogate for the PCA transformed dynamics back to the time-domain when constructing profile-likelihood based confidence intervals. This mapping is denoted by
\begin{equation}\label{eq:pce_PCA}
    \mathcal{M}^{\text{PCE}}\left(t;\boldsymbol{\theta}\right) = \boldsymbol{\mu}(\textbf{t}) + \sum_{q=1}^M \sum_{k=0}^{\numpcecoeff-1} \pcecoeff_{qk} \pcepoly_k(\boldsymbol{\param}) \eta_q(t)
\end{equation}
 where $\pcecoeff_{qk}$ is the PCE coefficient corresponding to the $q$th principal component and $k$th polynomial term, $\pcepoly_k(\boldsymbol{\param})$ is the $k$th polynomial basis function, and $\eta_q$ is the corresponding principal component basis vector, respectively. We build separate PCE-PCA emulators, $\mathcal{M}^{\text{PCE}}\left(t;\boldsymbol{\theta}\right)$ for each simulator output state used in our analysis, i.e., vessel specific pressure, flow, or area.

\subsection*{Global Sensitivity Analysis}
The use of spectral-surrogates enables the immediate calculation of variance-based, Sobol' indices once the surrogate is developed. We map the parameter space to the interval $[0,1]$ and denote the scaled parameter space by $\Gamma \in \mathbb{R}^{\numparam}$. We assume that the parameters of the system are independent in their prior space, which then enables a hierarchical decomposition of the model by
\begin{equation}
    \mathcal{M}(\param) = f_0 + \sum_{i=1}^{\numparam} f_i(\theta_i) + \sum_{1\leq i < j \leq \numparam} f_{ij}(\theta_{ij}) + \dots .
\end{equation}
where $ f_i(\theta_i)$ represents the contributions to $\mathcal{M}(\param)$ through the sole effects of $\theta_i$, $f_{ij}(\theta_{ij})$ represents the pairwise contributions, and so on. Under the assumption of independent priors on the parameters, the hierarchical components of the model are orthogonal, and enables a similar hierarchical decomposition of the variance of the system, $V = \var\left[\data\right]$. The partial-variances attributed to a given parameter are then expressed as
\begin{equation}
     V_i\left(\data\right) =\int_0^1 f^2_{i}(\theta_i) d\theta_i. 
\end{equation}
The first-order Sobol' index is defined by
\begin{equation}\label{eq:si}
    S_{i} = \frac{V_i}{V} = \frac{\var\left[\expect\left[\data | \theta_i\right]\right]}{\var\left[\data\right]}, .
\end{equation}
which represents the proportion of variance attributed to a single parameter, $\theta_i$, while the total-order Sobol' index is defined as
\begin{equation}\label{eq:st}
    S_{T_i} = 1 -  \frac{\var\left[\expect\left[\data | \bm{\theta}_{\sim i}\right]\right]}{\var\left[\data\right]}
\end{equation}
where the notation $\bm{\theta}_{\sim i}$ denotes the set of parameters that does not include $\theta_i$. The total-order index represents the proportion of variance attributed to the parameter $\theta_i$ through all of its interactions with other parameters, hence $S_{T_i}\approx 0$ indicates that $\theta_i$ has little contribution on the variance. This implies that it is functionally non-influential \cite{Nagel2020,Eck2016,Smith2024}.

We examine both the Sobol' indices for the PCA representation of the output, as well as for the time-dependent output following methods defined by Nagel et al. \cite{Nagel2020}. For the former, the values of $S^{q}_i$ and $S^{q}_{T_i}$, corresponding to the $q$th PCA score sensitivity, are given by
\begin{equation}
    S^{q}_i =   \frac{1}{\var\left[\bm{Z_q}\right]}\displaystyle \sum_{k\in\mathcal{A}_{S_i}}\left(\pcecoeff_{qk}^2\gamma_k\right), \ \ \
    S^{q}_{T_i} =  \frac{1}{\var\left[\bm{Z_q}\right]}\displaystyle\sum_{k\in\mathcal{A}_{S_{T_i}}}\left(\pcecoeff_{qk}^2\gamma_k\right),
\end{equation}
where $\mathcal{A}_i$ and $\mathcal{A}_{S_{T_i}}$ represent the set of polynomial coefficients corresponding to only $\theta_i$ and all coefficients that include $\theta_i$, respectively. The variable $\gamma_k$ represents the polynomial normalization factor defined previously. As described in detail by Nagel et al. \cite{Nagel2020}, the coefficients corresponding to the PCE-PCA surrogate can also be used to compute point-wise Sobol' indices describing the original, time-series output. Note that
\begin{eqnarray}
    &\expect\left[Y\mid \theta_i\right]\approx \expect\left[\mathcal{M}^{\text{PCE}}\mid \theta_i\right] \nonumber \\
    &= \boldsymbol{\mu}(\textbf{t}) + \displaystyle\sum_{q=1}^M \expect\left[c_q(\boldsymbol{\param})\mid \theta_i\right] \eta_q(t) \nonumber \\
    &= \boldsymbol{\mu}(\textbf{t}) + \displaystyle\sum_{q=1}^M \sum_{k=0}^{\numpcecoeff-1} \pcecoeff_{qk} \expect\left[\pcepoly_k(\boldsymbol{\param})\mid \theta_i\right] \eta_q
\end{eqnarray}
and similarly
\begin{equation}
    \expect\left[Y\mid \bm{\theta}_{\sim i}\right]\approx \boldsymbol{\mu}(\textbf{t}) + \displaystyle\sum_{q=1}^M \sum_{k=0}^{\numpcecoeff-1} \pcecoeff_{qk} \expect\left[\pcepoly_k(\boldsymbol{\param})\mid \bm{\theta}_{\sim i}\right] \eta_q
\end{equation}
where we remove the time-dependence of $\eta_q$ for ease of notation. We can then write
\begin{eqnarray}
    &\var\left[\expect\left[\data | \theta_i\right]\right] \approx \var\left[\expect\left[\mathcal{M}^{\text{PCE}} | \theta_i\right]\right] \nonumber \\
    = &\displaystyle\sum_{q=1}^M \var\left[\expect\left[c_q(\boldsymbol{\param})\mid \theta_i\right]\right] \eta^2_q + 2\sum_{q<q^*}\textbf{Cov}\left[\expect\left[c_q(\boldsymbol{\param})\mid \theta_i\right],\expect\left[c_{q^*}(\boldsymbol{\param})\mid \theta_i\right]\right]\eta_q\eta_{q^*} \nonumber \\
    = &\displaystyle\sum_{q=1}^M S_i^q \var\left[c_q(\bm{\theta})\right] \eta^2_q + 2\sum_{q<q^*}\textbf{Cov}\left[\expect\left[c_q(\boldsymbol{\param})\mid \theta_i\right],\expect\left[c_{q^*}(\boldsymbol{\param})\mid \theta_i\right]\right]\eta_q\eta_{q^*}
\end{eqnarray}
where on the last line we rearrange the definition of $S_i^q$ provided in eq \eqref{eq:si}. By similar argument, we can rearrange the definition of $S_{T_i}^q$ and write
\begin{eqnarray}
     &\var\left[\expect\left[\data | \bm{\theta}_{\sim i}\right]\right] \approx \var\left[\expect\left[\mathcal{M}^{\text{PCE}} | \bm{\theta}_{\sim i} \right]\right] \nonumber \\
    = &\displaystyle\sum_{q=1}^M \left(1-S_{T_i}^q\right) \var\left[c_q(\bm{\theta})\right] \eta^2_q + 2\sum_{q<q^*}\textbf{Cov}\left[\expect\left[c_q(\boldsymbol{\param})\mid \bm{\theta}_{\sim i}\right],\expect\left[c_{q^*}(\boldsymbol{\param})\mid \bm{\theta}_{\sim i}\right]\right]\eta_q\eta_{q^*}. \ \ \ \ 
\end{eqnarray}
The point-wise first and total-order Sobol' indices are then given by
\begin{eqnarray}\label{eq:sob_time}
    S_i^Y = \displaystyle \frac{\displaystyle\sum_{q=1}^M S_i^q \var\left[c_q(\bm{\theta})\right] \eta^2_q    + 2\sum_{q<q^*}\textbf{Cov}\left[\expect\left[c_q(\boldsymbol{\param})\mid \theta_i\right],\expect\left[c_{q^*}(\boldsymbol{\param})\mid \theta_i\right]\right]\eta_q\eta_{q^*}}{{\var\left[\data\right]}}
\end{eqnarray}
\begin{eqnarray}
    S_{T_i}^Y =  1 - \displaystyle \frac{\displaystyle\sum_{q=1}^M \left(1-S_{T_i}^q\right) \var\left[c_q(\bm{\theta})\right] \eta^2_q + 2\sum_{q<q^*}\textbf{Cov}\left[\expect\left[c_q(\boldsymbol{\param})\mid \bm{\theta}_{\sim i}\right],\expect\left[c_{q^*}(\boldsymbol{\param})\mid \bm{\theta}_{\sim i}\right]\right]\eta_q\eta_{q^*}}{{\var\left[\data\right]}}.
\end{eqnarray}
As noted in the original derivation of this approach by Nagel et al. \cite{Nagel2020}, the orthogonality of the polynomials provides the following covariance definitions:
\begin{eqnarray}
    \textbf{Cov}\left[\expect\left[c_q(\boldsymbol{\param})\mid \theta_i\right],\expect\left[c_{q^*}(\boldsymbol{\param})\mid \theta_i\right]\right] = \displaystyle \sum_{k\in\mathcal{A}_{S_i}}z_{qk}z_{q^{^*} k} \\
    \textbf{Cov}\left[\expect\left[c_q(\boldsymbol{\param})\mid \bm{\theta}_{\sim i}\right],\expect\left[c_{q^*}(\boldsymbol{\param})\mid \bm{\theta}_{\sim i}\right]\right] = \displaystyle \sum_{k\in\mathcal{A}_{S_{T_i}}}z_{qk}z_{q^{^*} k} 
\end{eqnarray}

\subsection*{Profile-likelihood Confidence Intervals}\label{sec:PL}
The profile-likelihood is a frequentist statistical method that explicitly calculates parameter confidence intervals by ``profiling" a single parameter at a time \cite{Wieland2021}. Let $\mathcal{M}(\theta)$ denote the model output and $\mathbf{y}$ be the corresponding, possibly noisy, observations. The negative log-likelihood is then proportional to the weighted sum of square error
\begin{equation}\label{eq:loglike}
    -LL(\bm{\theta}) \propto \left(\mathbf{y} -\mathcal{M}(\bm{\theta}) \right)^\top \mathbf{\Sigma^{-1}}\left(\mathbf{y} -\mathcal{M}(\bm{\theta}) \right).
\end{equation}
The weight matrix, $\mathbf{\Sigma}$, has diagonal entries that represent the variance of the measurement noise and off-diagonal values representing the covariances between observations. Here, we assume noise free, independent measurements and assign the diagonal elements of $\mathbf{\Sigma}$ to a scalar value that nondimensionalizes different observations used in the experimental design. Specifically, we consider three experimental designs: (i) only dynamic pressure data in the first branch, (ii) dynamic pressure in the first branch and dynamic area in the daughter branches, and (iii) dynamic pressure in the first branch and dynamic flow in the daughter branches. In these cases, the diagonal elements of $\mathbf{\Sigma}$ are the average pressure, flow, or area values of the data.

\begin{figure}
    \includegraphics[width=\linewidth]{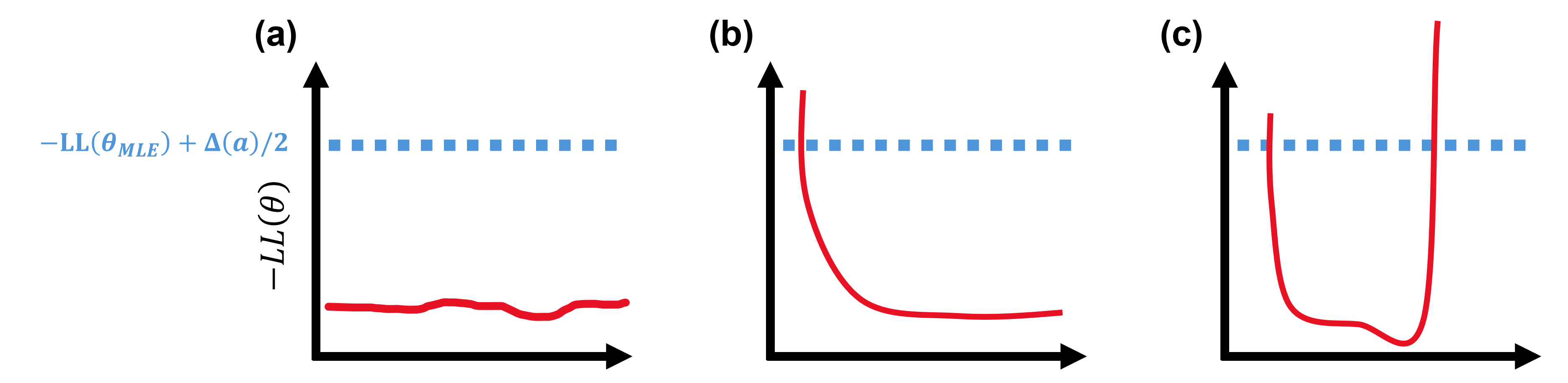}
    \caption{Example one-dimensional profile-likelihood results and upper bounds for identifiability classification. An example identifiability cutoff given by the maximum likelihood estimate, $-LL(\bm{\theta}_{MLE})$, and chi-squared statistics, $\Delta(a)/2$ are provided in blue. A flat profile-likelihood (a) suggests structural or practical identifiability issues, whereas a profile-likelihood that is bounded on one side (b) by the likelihood threshold is considered practically non-identifiable. In contrast, a profile-likelihood that is completely bounded (c) implies a practically identifiable parameter.}
    \label{fig:PL_Example}
\end{figure}

The profile-likelihood for a parameter $\theta_i$ is then defined as
\begin{equation}\label{eq:PL}
    PL_i(\theta_i) = \min_{\bm{\param}_{\sim i}} \left[-LL\left(\bm{\param}_{\sim i} | \theta_i, \mathbf{y}\right)\right]
\end{equation}
where $\theta_i$ is the parameter being profiled and $\bm{\param}_{\sim i}=\bm{\theta} \setminus \{\theta_i\}$ is the remaining parameters to be inferred (similar to the definition in Sobol' indices). Univariate confidence intervals can be computed for each parameter by comparing the profile-likelihood to a chi-square distribution
\begin{equation}
    \text{CI}(\theta_i) = \left\{\theta_i | -2\left(PL_i(\theta_i) - LL(\bm{\theta}_{MLE}|\mathbf{y})\right) \leq \Delta(a)\right\}
\end{equation}
where $\bm{\theta}_{MLE}$ is the maximum likelihood estimator (MLE) of the full parameter set corresponding to the maximum likelihood (minimum of the negative log-likelihood). The difference between the profile-likelihood and MLE evaluated log-likelihood are compared to a chi-squared distribution, $\Delta(a)=\text{icdf}\left(\chi^2_1,1-a\right)$, with one-degree of freedom and an $1-a$ confidence level. Prototypical profile-likelihood plots can be found in Figure \ref{fig:PL_Example}. Parameter confidence intervals that are flat suggest non-identifiable parameters, as there are infinite confidence bounds for which the parameter value may take on (Figure \ref{fig:PL_Example}(a)). This may be a practical or a structural identifiability issue \cite{Wieland2021,Boiger2016,Colebank2022}. A confidence interval that is bounded on one side alone suggests practical identifiability issues, as more data in the design or a reduction in measurement noise may lead to finite confidence bounds on both sides of the MLE (Figure \ref{fig:PL_Example}(b)). Finally, we consider a parameter practically identifiable if there exist finite confidence bounds around the MLE, providing a typical parabolic negative log-likelihood with finite confidence bounds (Figure \ref{fig:PL_Example}(c)).

Here, we consider the spectral, PCA emulator, $\mathcal{M}^{\text{PCE}}\left(t;\boldsymbol{\theta}\right)$ defined in eq \eqref{eq:pce_PCA} for calculated the negative log-likelihood. We note the emulator based negative log-likelihood by $-\widetilde{LL}(\bm{\theta})$. We consider point-wise confidence intervals at a 95\% significance level throughout. As noted previously, we build an independent PCE-PCA emulator for each output quantity stemming from the PDE simulator. In this work, we are interested in understanding how different experimental designs affect practical identifiability. Thus, using the PCE-PCA emulator and the profile-likelihood confidence intervals, we assess three different experimental designs, $Di$, for the pulmonary circulation model:
\begin{enumerate}
    \item $D1$: inference using only pressure time-series data in the MPA;
    \item $D2$: inference using pressure time-series data in the MPA and time-series area data in the LPA and RPA; and
    \item $D3$: inference using pressure time-series data in the MPA and time-series flow data in the LPA and RPA.
\end{enumerate}
We set the measurement noise covariance in eq \eqref{eq:loglike} to be a diagonal matrix with entries given by the average of the data in the experimental design, i.e. $\bm{\Sigma}=\text{diag}\left(\bar{\bm{y}}_{Di}\right)$, where $\bar{\bm{y}}_{Di}$ is a vector with the average of each data source included in experimental design $Di$. This scaling of the components in the likelihood is crucial, as the model outputs vary in both units and magnitude, and will bias the profile-likelihood construction if not properly implemented. We note that this definition of $\bm{\Sigma}$ is specific to our application, and would instead represent the measurement error variances in a typical multioutput inverse problem \cite{Colebank2022}.

When calculating the profile-likelihood, we always begin by inferring the full set of parameters. As shown in the results, we subsequently analyze the profile-likelihood for smaller parameter subsets to investigate parameter identifiability. This reflects a more realistic scenario than fixing parameters at their true value prior to inference, as we the true data generating parameters are unknown.

\section*{Results}
\subsection*{Emulator Accuracy on Test Data}

We first assess the accuracy of our PCE emulator on a set of test data for both boundary conditions. We use a degree five polynomial for the PCE emulator, which has been shown previously to achieve excellent accuracy in emulating the reduced dimensional output \cite{Paun2025}. We use the mean square relative error (MSRE) as a metric for emulator accuracy, given by
\begin{equation}
    \text{MSRE} = \frac{1}{N_{samp}}\sum_{i=1}^{N_{samp}} \left(\frac{\bm{Y}\left(\param^i\right)-\mathcal{M}^{\text{PCE}}(\param^i)}{\max \left(\bm{Y}\left(\param^i\right)\right)}\right)^2.
\end{equation}

\begin{figure}[h!]
    
        \begin{subfigure}[c]{\textwidth}
            \centering
            \includegraphics[width=0.6\linewidth]{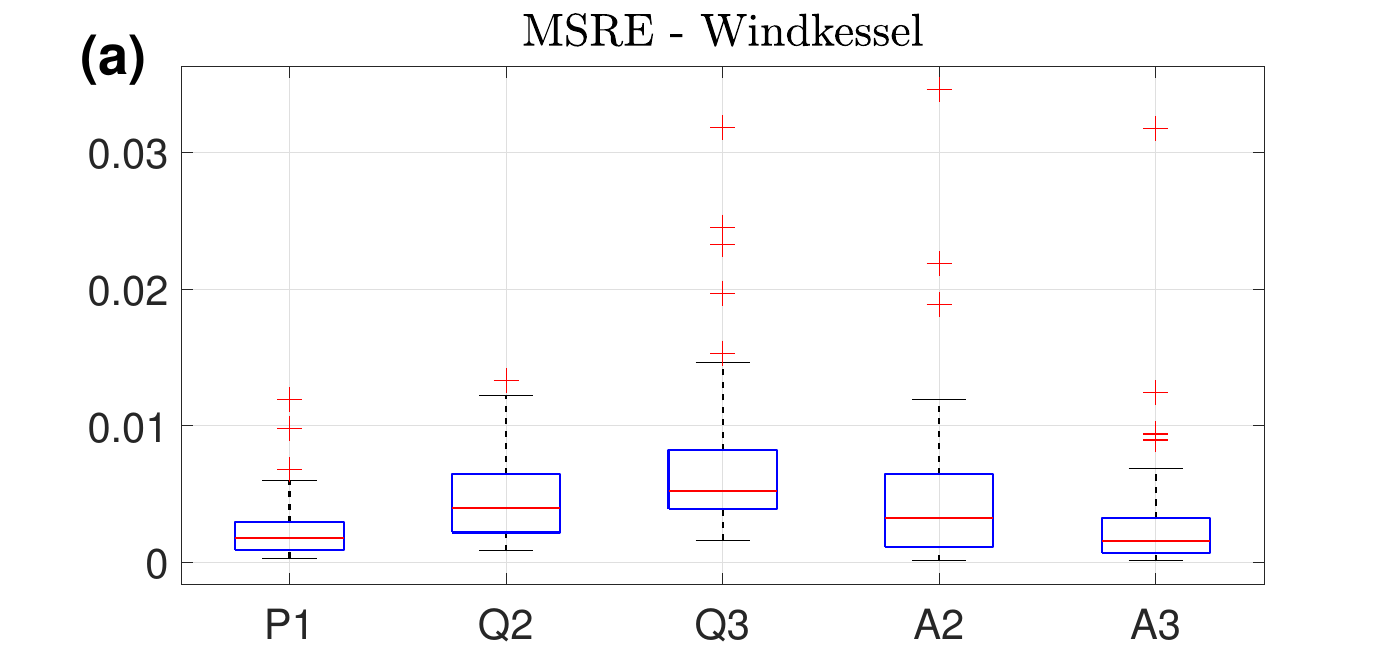}
        \end{subfigure}
        \begin{subfigure}[c]{\textwidth}
            \centering
            \includegraphics[width=0.6\linewidth]{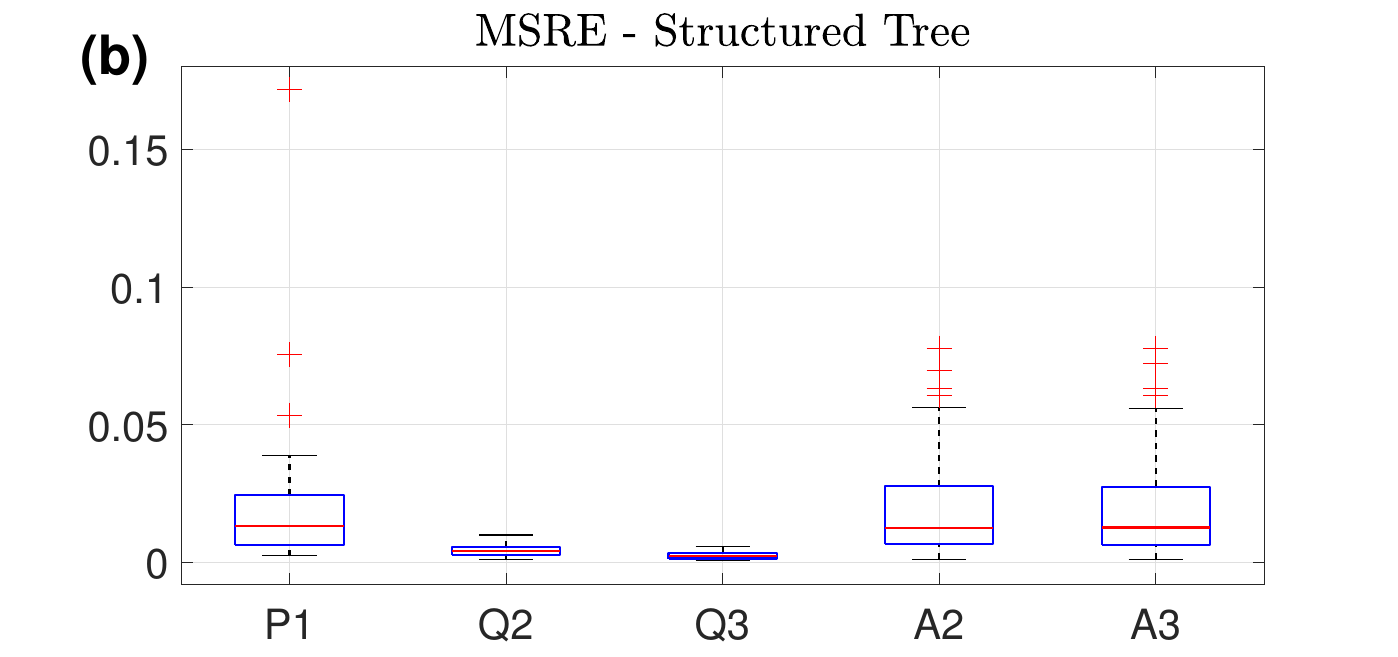}
        \end{subfigure}
    \caption{Emulator accuracy on 50 test data sets for pressure in the MPA (P1), flows in the LPA and RPA (Q2 and Q3), and areas in the LPA and RPA (A2 and A3). (a) Windkessel boundary conditions. (b) Structured tree boundary conditions.}
    \label{fig:testdata}
\end{figure}

The model is built on the first five principal components of each vessel's pressure, flow, or area output corresponding to the designs $D1,D2,$ and $D3$. Table \ref{tab:Variance} shows the proportion of variance attributed to each component. We use a degree five PCE polynomial for both sets of boundary condition models. As shown in Figure \ref{fig:testdata}, the PCE-PCA accuracy is relatively high in the Windkessel model, with a median MSRE on the order of 2\% or less for all five QoIs. There is no drastic difference in accuracy, although pressure emulation tends to have the lowest median MSRE in the Windkessel model. In contrast, the structured tree model shows relatively larger MSRE values. The pressure MSRE is largest, with a median value of $\approx 2\%$ and outliers at 9\% and 15\%. Area error metrics are also larger, though the median value is still $\approx 2\%$. The flow errors are relatively similar in magnitude of those from the Windkessel model.

\begin{table}[h]
    \centering
    \begin{tabular}{lcccccc}
        \hline
        QoI & PC 1 (\%) & PC 2 (\%) & PC 3 (\%) & PC 4 (\%) & PC 5 (\%) & Variance (\%) \\
        \hline
        \multicolumn{7}{c}{\textbf{Windkessel Model}}\\
        \hline
        Ves 1 P & 82.2 & 16.4 & 1.4 & 0.0 & 0.0 & 99.9 \\
        \hline
        Ves 2 Q & 58.2 & 34.3 & 7.1 & 0.3 & 0.0 & 99.9 \\
        Ves 2 A & 99.9 & 0.1 & 0.0 & 0.0 & 0.0 & 100.0 \\
        \hline
        Ves 3 Q & 58.1 & 33.6 & 6.2 & 0.4 & 0.3 & 98.7 \\
        Ves 3 A & 99.9 & 0.1 & 0.0 & 0.0 & 0.0 & 100.0 \\
        \hline 
        \multicolumn{7}{c}{\textbf{Structured Tree Model}}\\
        \hline
        Ves 1 P & 88.2 & 9.8 & 1.9 & 0.1 & 0.0 & 99.9 \\
        \hline
        Ves 2 Q & 65.8 & 18.1 & 6.5 & 5.1 & 1.6 & 97.0 \\
        Ves 2 A & 97.9 & 1.8 & 0.3 & 0.0 & 0.0 & 100.0 \\
        \hline
        Ves 3 Q & 64.3 & 14.8 & 9.8 & 4.6 & 2.3 & 95.7 \\
        Ves 3 A & 97.9 & 1.8 & 0.3 & 0.0 & 0.0 & 100.0 \\
        \hline
    \end{tabular}
    \caption{Variance attributed to the first five principal components (PCs). Right most column represents cumulative total variance from the five principal components (\%).}
    \label{tab:Variance}
\end{table}

\subsection*{Global sensitivity analysis}
\begin{figure}[h!]
    \centering
    \begin{subfigure}[b]{\textwidth}
        \includegraphics[width=0.5\linewidth]{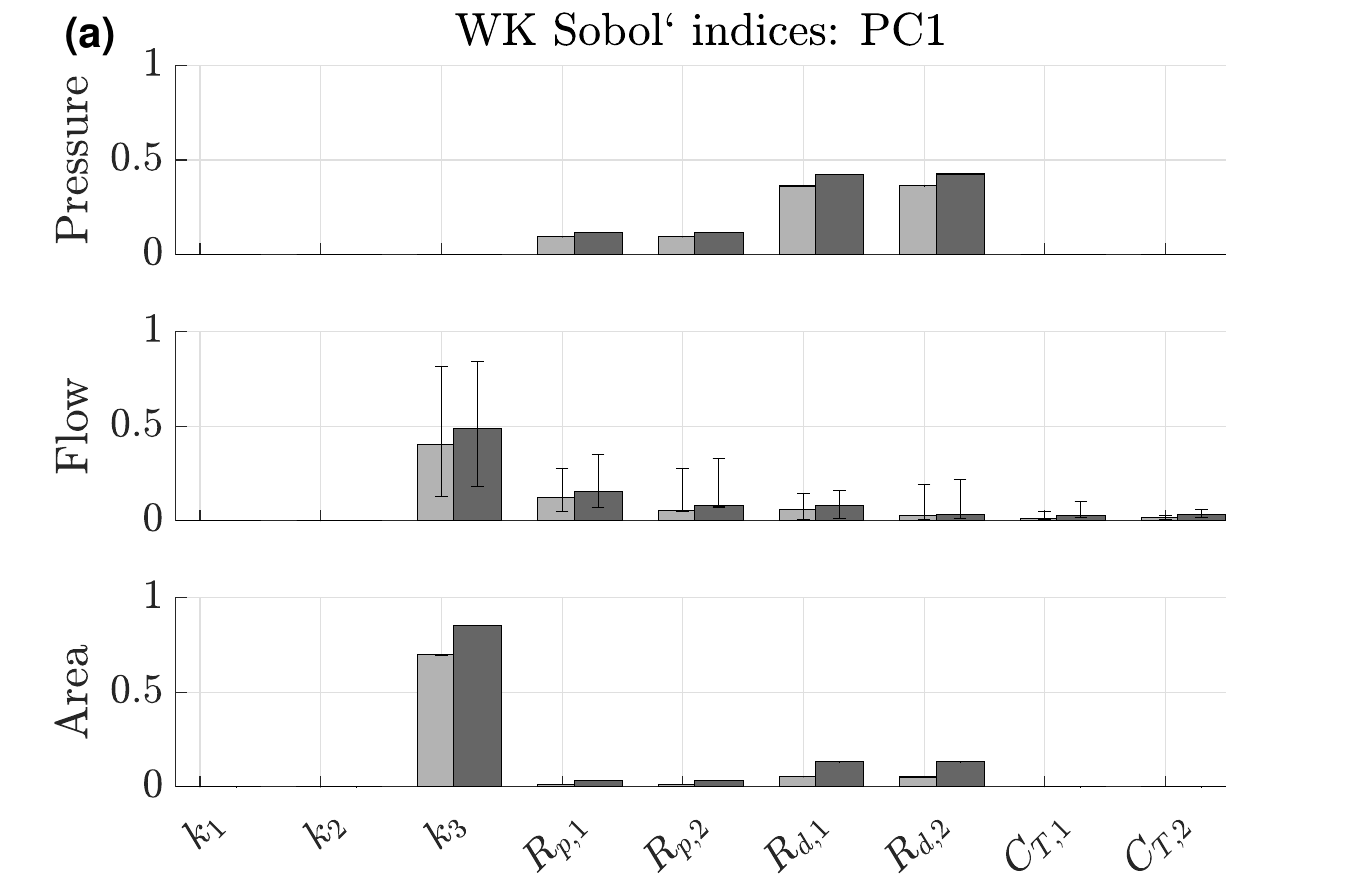}
        \includegraphics[width=0.5\linewidth]{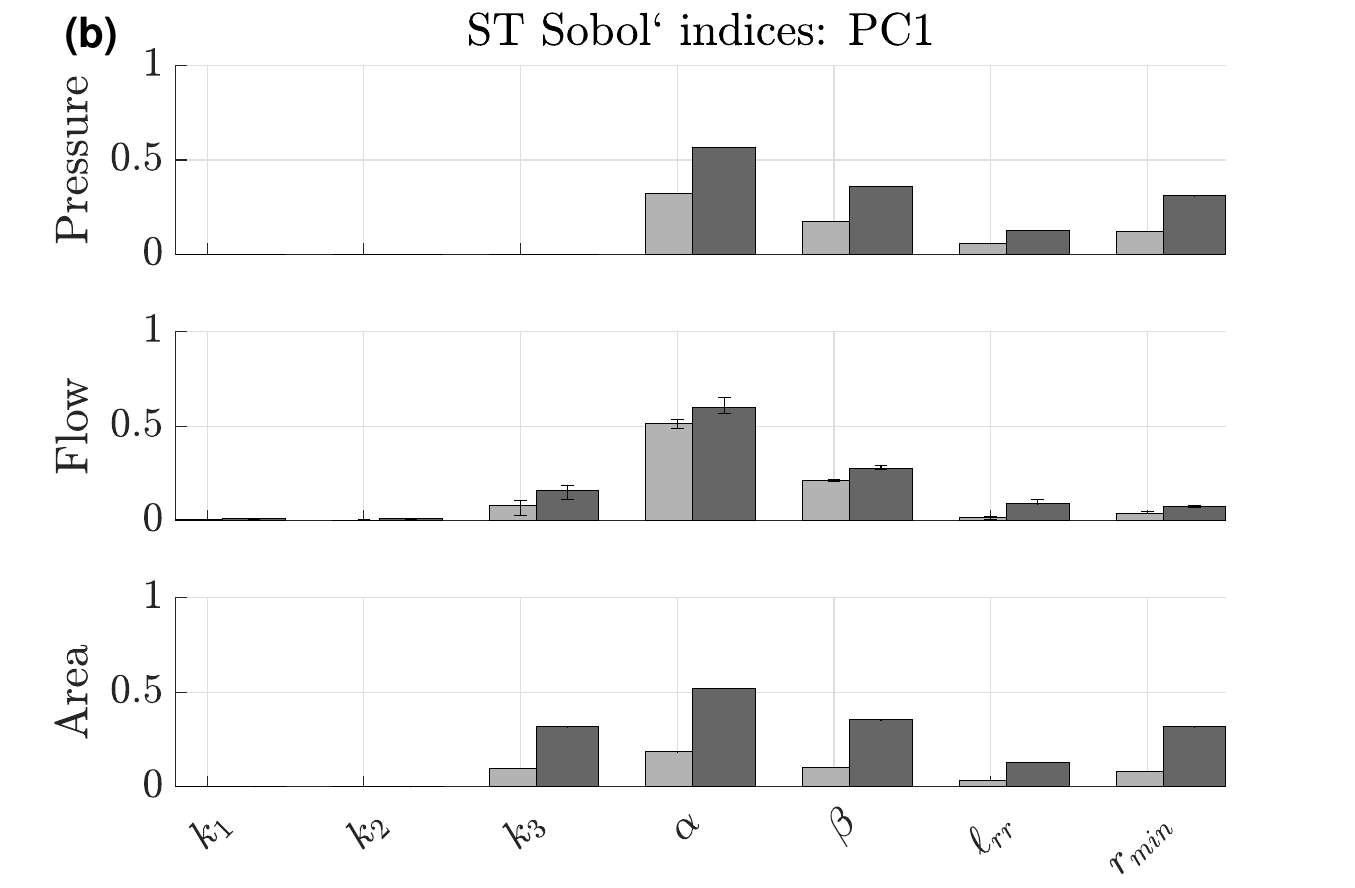}
        \includegraphics[width=0.5\linewidth]{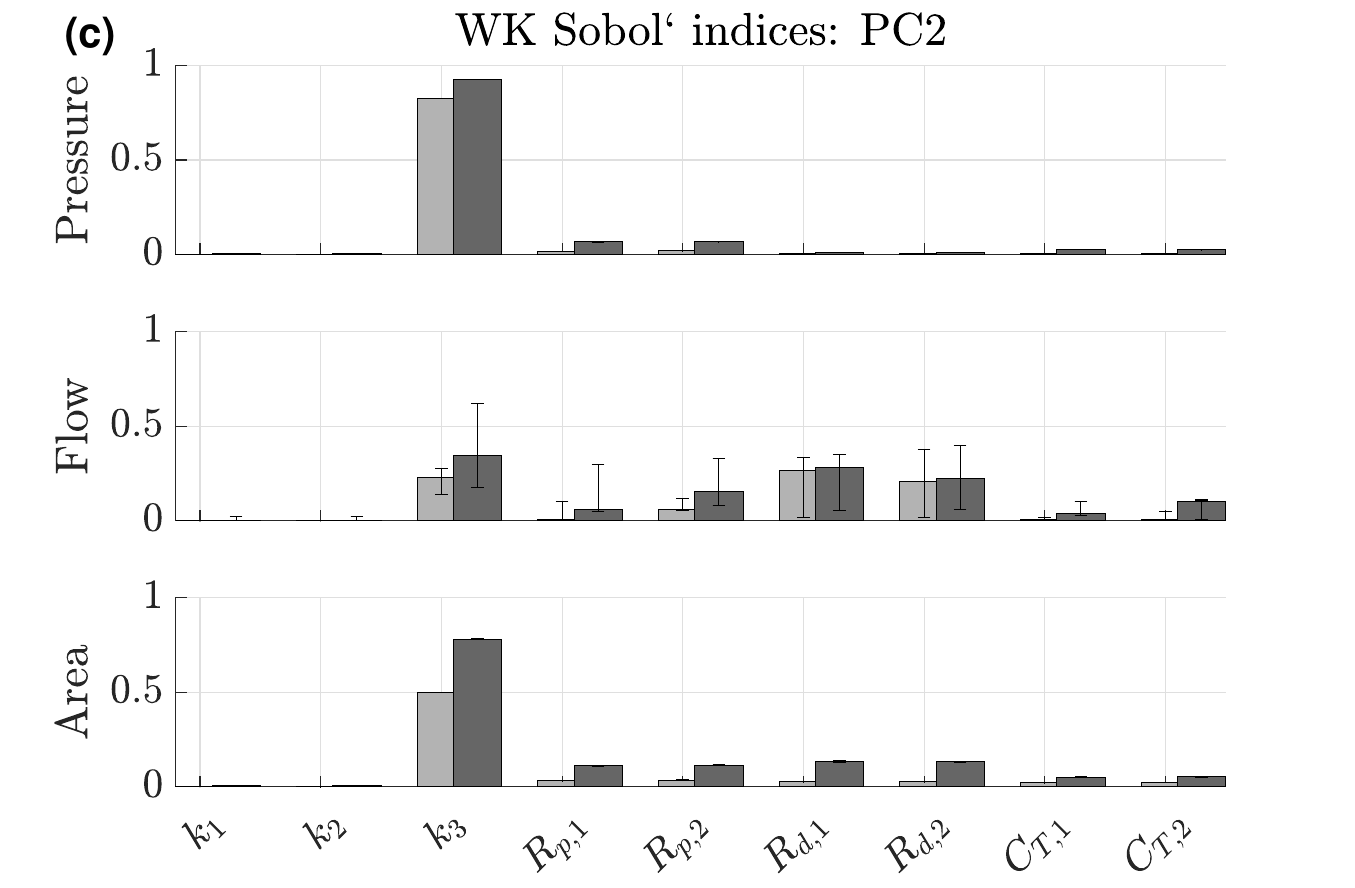}
        \includegraphics[width=0.5\linewidth]{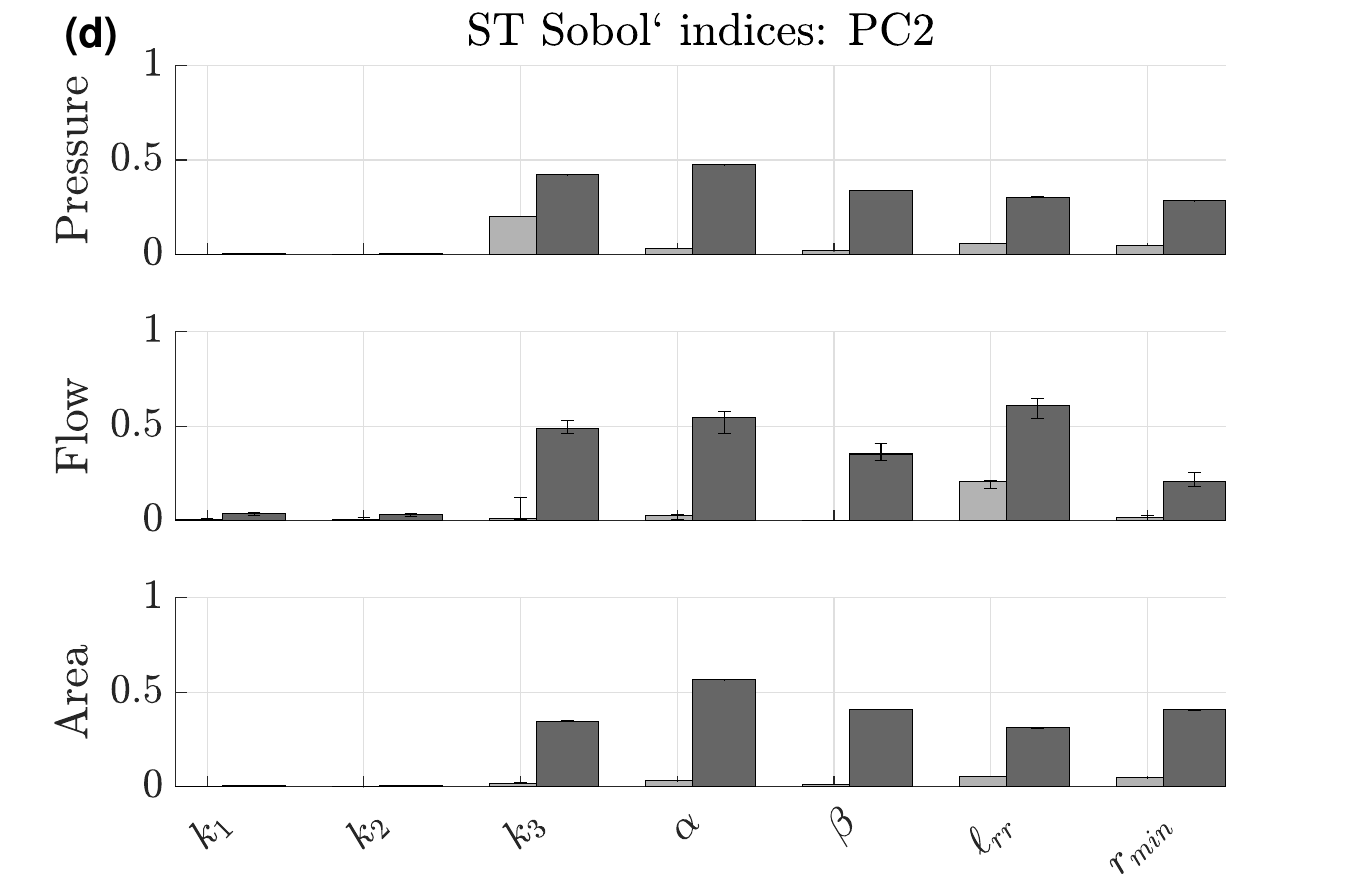}
        \includegraphics[width=0.5\linewidth]{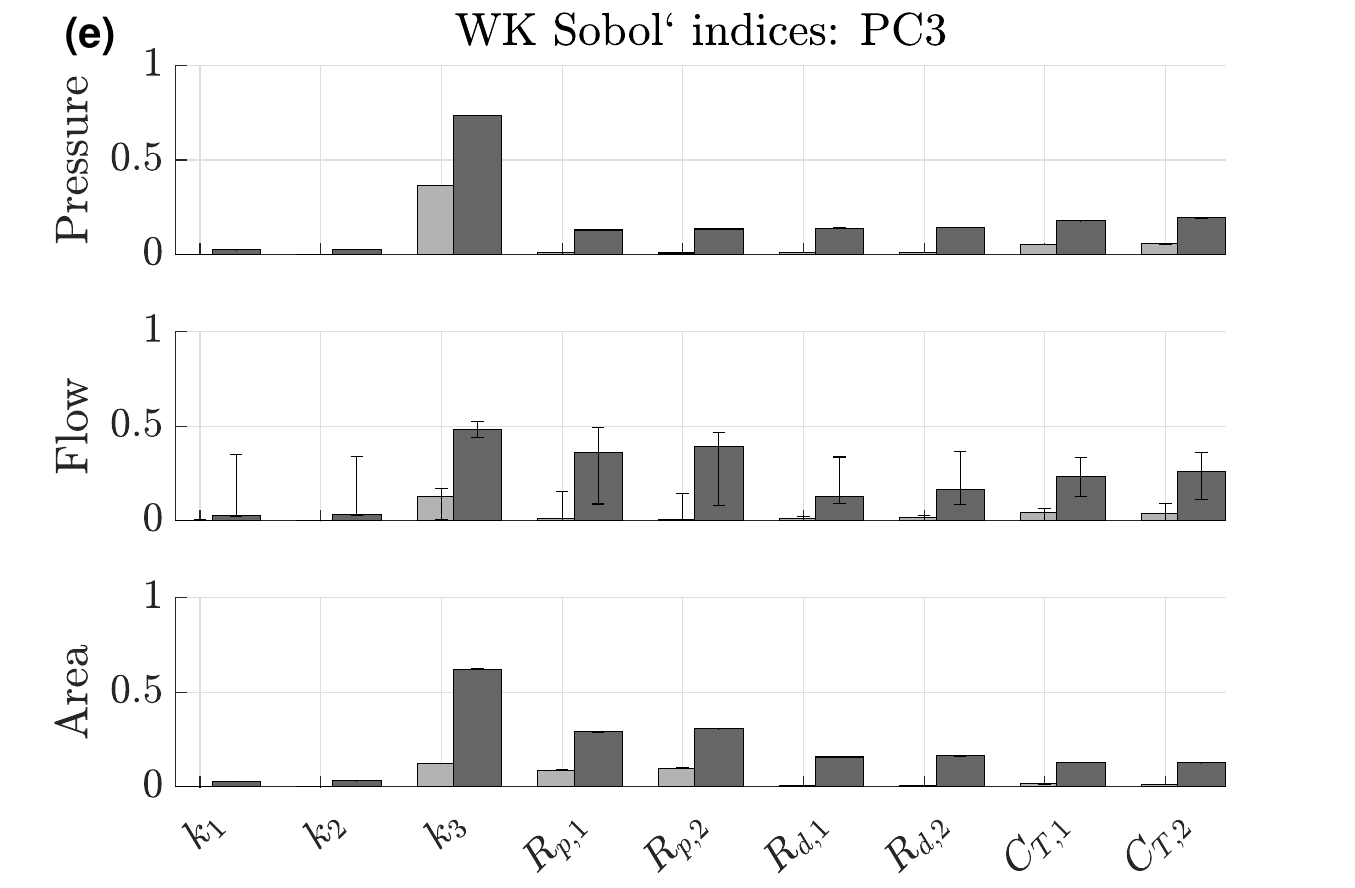}
        \includegraphics[width=0.5\linewidth]{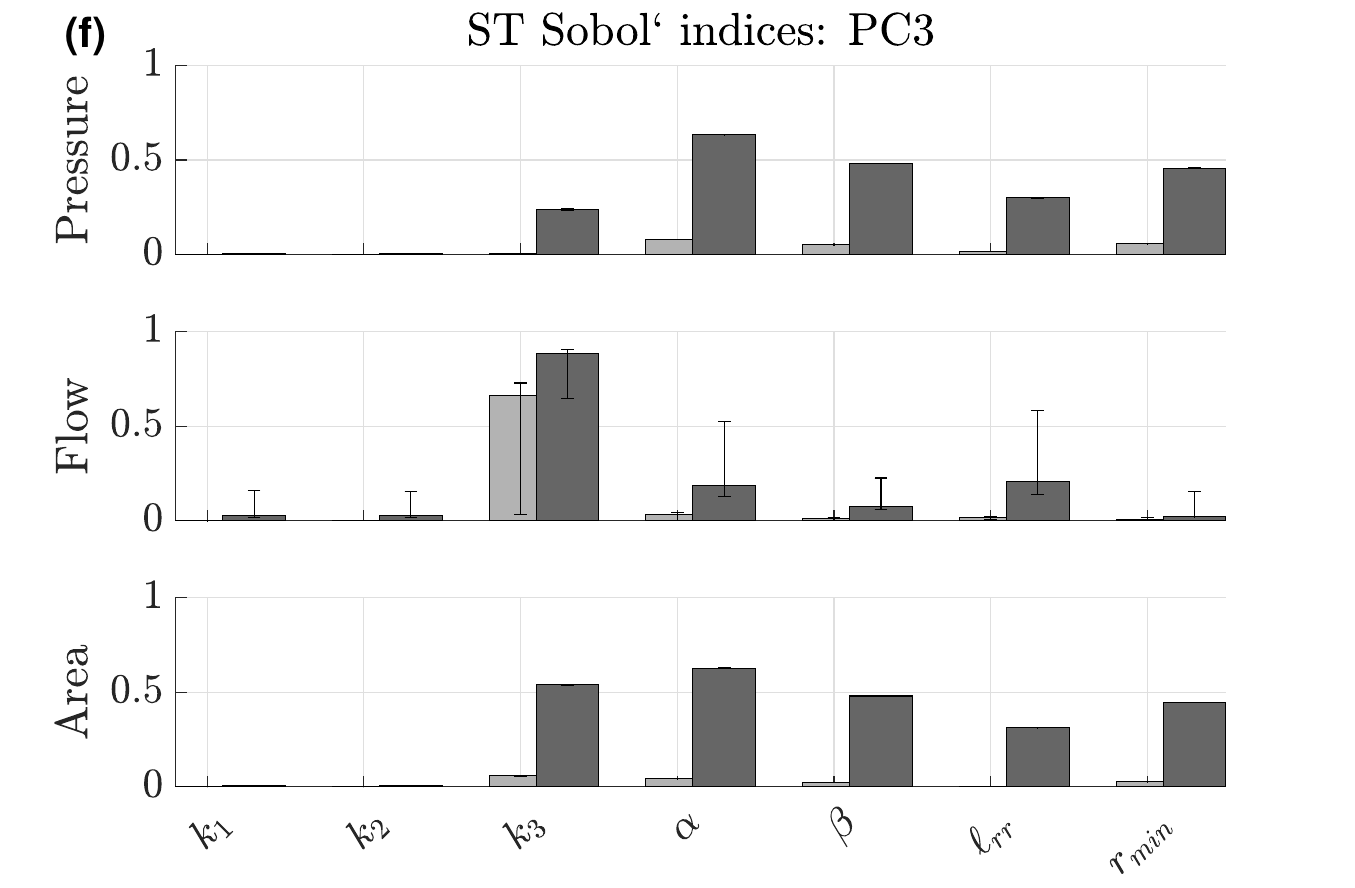}
        \end{subfigure}
    \caption{Sobol' indices for the first three principal components (PCs) using Windkessel boundary conditions (a,c,e) and structured tree boundary conditions (b,d,f). Light bars indicate the median first-order index, $S^q_{i}$, across the three vessels while dark grey represent the median total-order index. Error bars denote the range for the metrics across the three blood vessels.}
    \label{fig:Sobol_PCA}
\end{figure}

We use Sobol' indices (PCA based and pointwise) to determine how influential parameters are on each output. Figure \ref{fig:Sobol_PCA} shows the Sobol' indices across the first three principal components for each boundary condition, with bar graphs and error bars showing the median and range of Sobol' values across the MPA, LPA, and RPA. Results for the Windkessel boundary conditions are provided in Figure \ref{fig:Sobol_PCA}(a,c,e). The distal resistances $R_{d,1}$ and $R_{d,2}$ are most influential on the first principal component of pressure, whereas $k_3$ tends to dominate model sensitivity for flow and area. The proximal resistors $R_{p,1}$ and $R_{p,2}$ have relatively small effect on the first principal component of pressure and area, with some variable effects on flow. The exponential stiffness terms $k_1$ and $k_2$ are the least influential parameters for the first principal component, with the compliance parameters being minimally influentially as well. The second and third principal components are more sensitive to $k_3$, $R_{p,1}$ and $R_{p,2}$, with the high principal components showing elevated $S^q_{T_i}$ values for previously non-influential parameters. In general, the higher the principal component (and smaller the unexplained variance of the true simulator), the more uniform the $S^q_{T_i}$ values are, with first order indices $S_{i}$ shrinking in magnitude.

\begin{figure}[h!]
    \centering
    \begin{subfigure}[b]{\textwidth}
         \centering
        \includegraphics[width=0.9\linewidth]{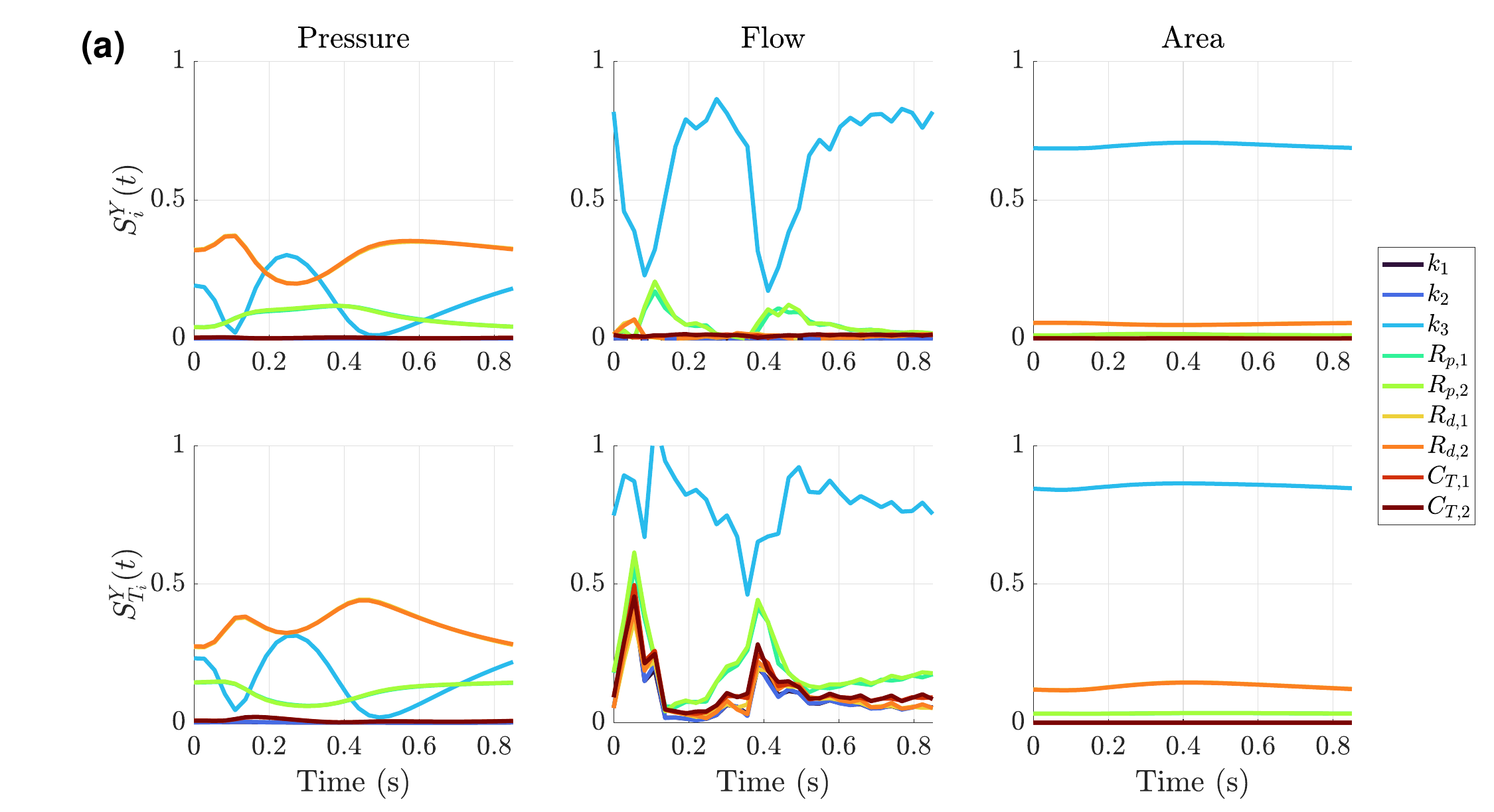}\vspace{-4mm}
        \includegraphics[width=0.9\linewidth]{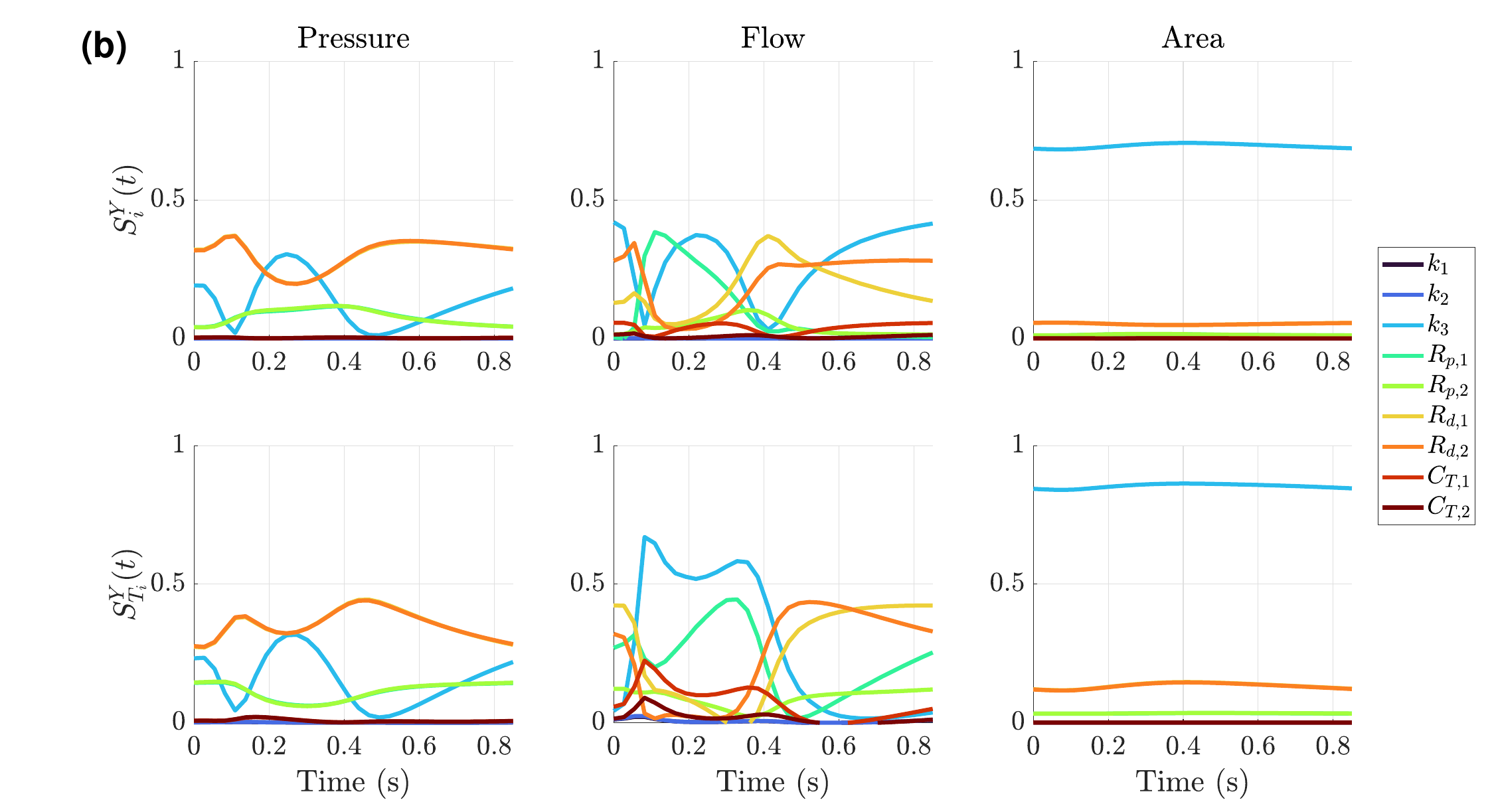}\vspace{-4mm}
        \includegraphics[width=0.9\linewidth]{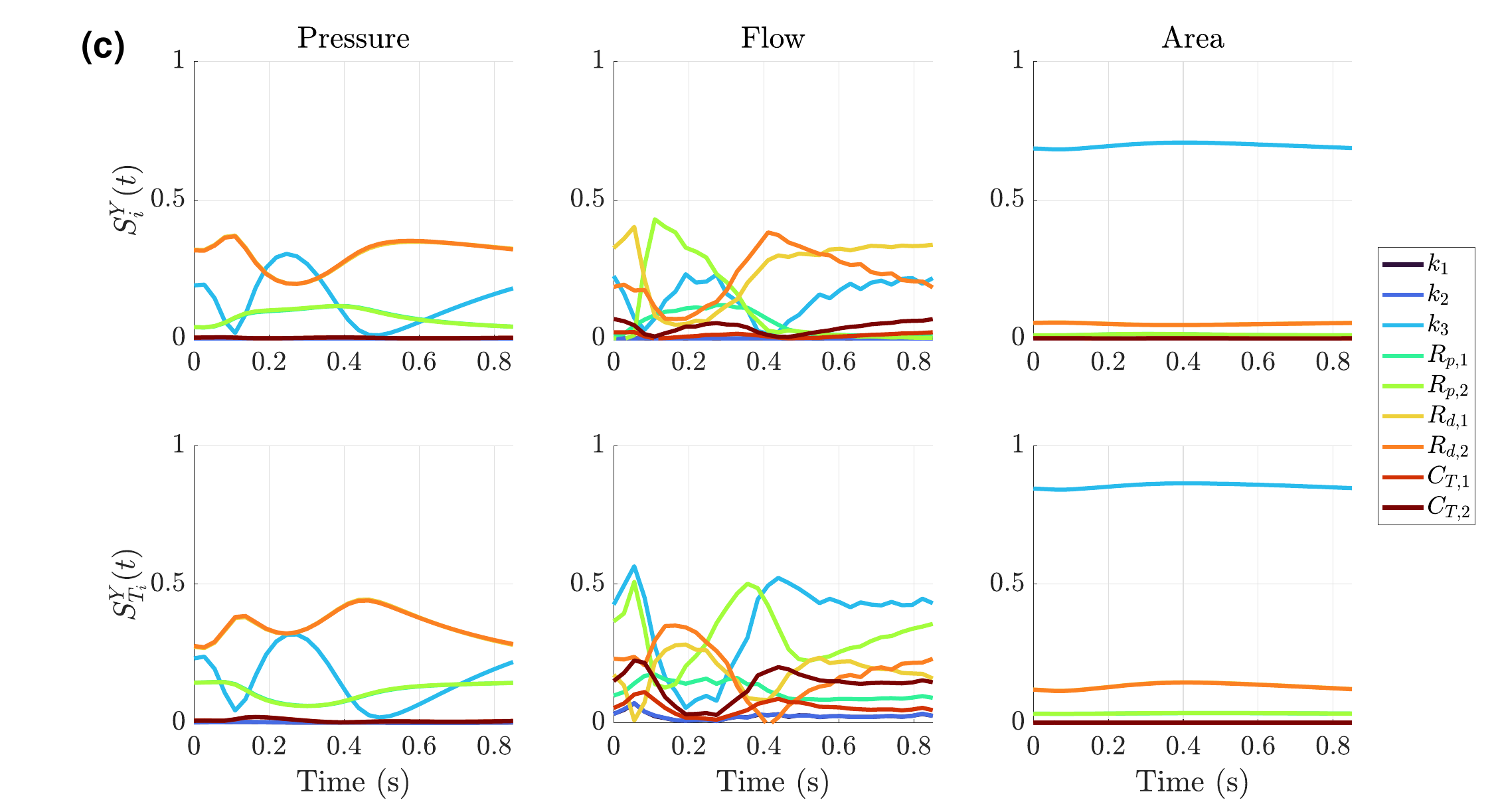}
    \end{subfigure}
    
    \caption{Point-wise Sobol' indices for the Windkessel boundary conditions calculated from principal component Sobol' metrics $S_i^{q}$ and $S_{T_i}^{q}$ via eq \eqref{eq:sob_time}. Results are presented for the (a) MPA, (b) LPA, and (c) RPA in the pulmonary tree.}
    \label{fig:Sobol_WK_time}
\end{figure}

We compare these findings to the reconstructed point-wise estimates, $S_i^{Y}(t)$ and $S_{T_i}^{Y}(t)$ in Figure \ref{fig:Sobol_WK_time} built using eq \eqref{eq:sob_time}. Results are shown for each vessel. Similar to the findings in Figure \ref{fig:Sobol_PCA}(a,c,e), the parameters $R_{d,1}$ and $R_{d,2}$ dominant the pressure sensitivity (they overlap exactly), with $k_3$ showing the next leading influence on the model for both  $S_i^{Y}(t)$ and $S_{T_i}^{Y}(t)$ metrics across all three branches. The $R_{p,1}$ and $R_{p,2}$ parameters are then the next most influential (again overlapping in magnitude and behavior), in some instances having higher sensitivity metrics than $k_3$. The effects of $k_1$, $k_2$, $C_{T,1}$, and $C_{T,2}$ are negligible in comparison. The point-wise sensitivity of the flow outputs are distinct across the three branches. The MPA flow output is located downstream from the prescribed flow boundary condition, and thus has a relatively small variance compared to the other two branches. The first-order sensitivities show that $k_3$ is by far the most influential. A similar finding is true for the total-order index, through the other parameters of the system appear to contribute equally. We note that the values of $S_{T_i}^{Y}(t)$ exceed 1.0 in the MPA, which is attributed to a numerical approximation error in the principal component decomposition (note that this issue does not arise when using four principal components only). For the LPA, we see that flow is most sensitive to $k_3$, $R_{p,1}$ during systole (0.1 - 0.3 seconds) and then most sensitive to $k_3$, $R_{d,1}$, and $R_{d,2}$ during diastole. The total-indices reflect similar trends, with the compliance term $C_{T,1}$ also contributing during the upstroke in flow. The RPA flow has elevated sensitivity to $k_3$, $R_{p,2}$, and both distal resistances $R_{d,1}$ and $R_{d,2}$, in terms of both $S_i^Y(t)$ and $S_{T_i}^{Y}(t)$. Again, the compliance to this branch, $C_{T,2}$ now appears moderately influential as was the case for the LPA. Finally, the area sensitivity is dominated by $k_3$, with only the distal resistance parameters $R_{d,1}$, and $R_{d,2}$ having some effects on area dynamics. This is consistent with the results in Figure \ref{fig:Sobol_PCA}(a,c), which shows $k_3$ dominating across the first two principal components for all branches.


The structured tree model results in Figure \ref{fig:Sobol_PCA}(b,d,f) show similar minimal influence by $k_1$ and $k_2$ across all outputs for the first three principal components as the Windkessel model. Pressure is most sensitive to the structured tree parameters $\alpha,\beta,\ell_{rr},$ and $r_{min}$ for the first principal component. The second and third pressure principal components are increasingly more affected by $k_3$. In general $\alpha$ and $\beta$ appear most influential, following by $k_3$, $r_{min}$, and $\ell_{rr}$. The flow results also indicate that the five parameters excluding $k_1$ and $k_2$ dominate model sensitivity. The first through third principal components show that $alpha$, $\beta$, $k_3$, and $\ell_{rr}$ drive flow sensitivity. There tends to be large variability between branches for the third, fourth, and fifth principal component (latter two not shown). Finally, the area predictions are similar in sensitivity to pressure, with the exception that $k_3$ is more influential on area. 

\begin{figure}[h!]
    \centering
    \begin{subfigure}[b]{\textwidth}
         \centering
        \includegraphics[width=0.9\linewidth]{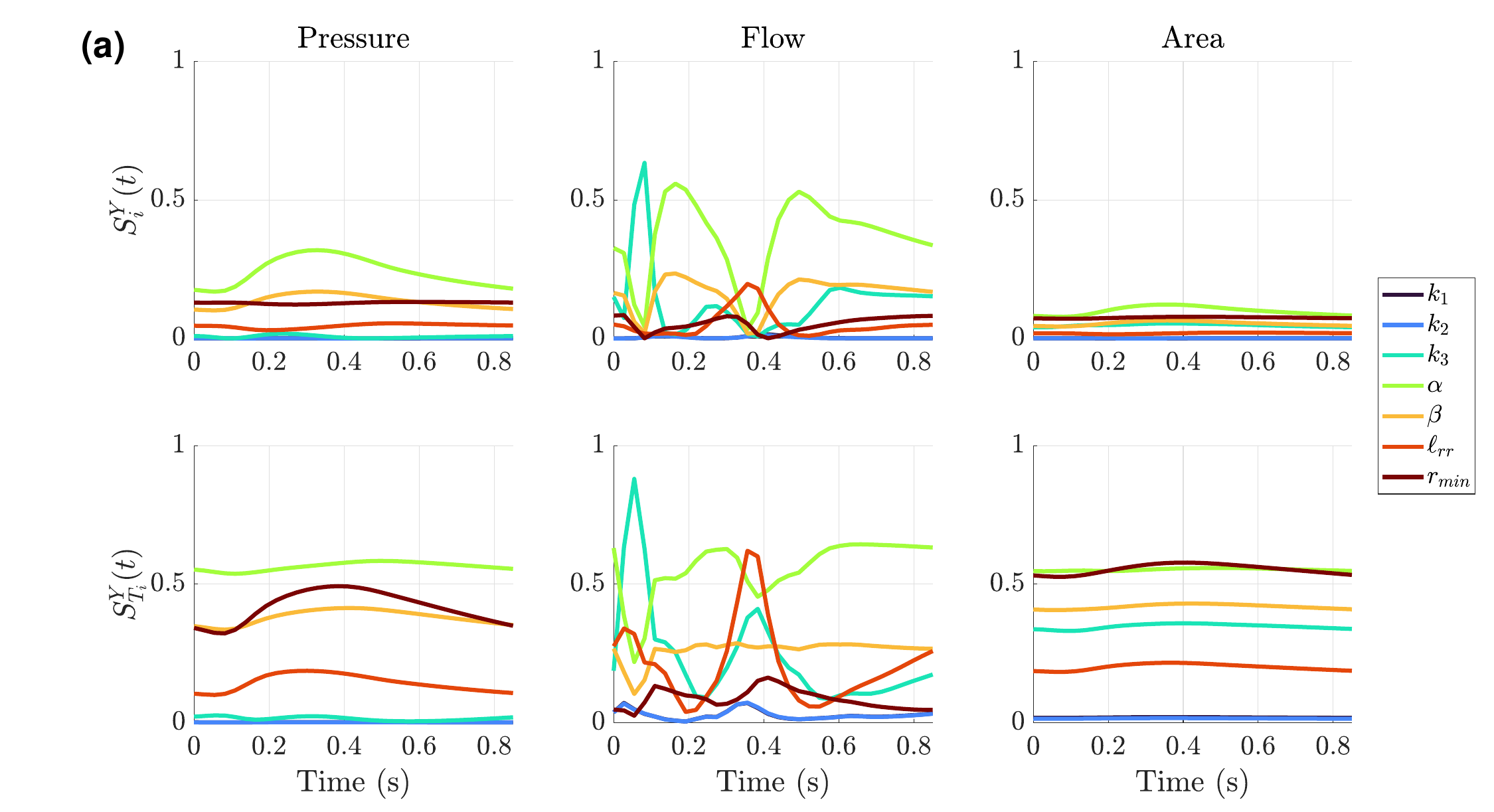}\vspace{-4mm}
        \includegraphics[width=0.9\linewidth]{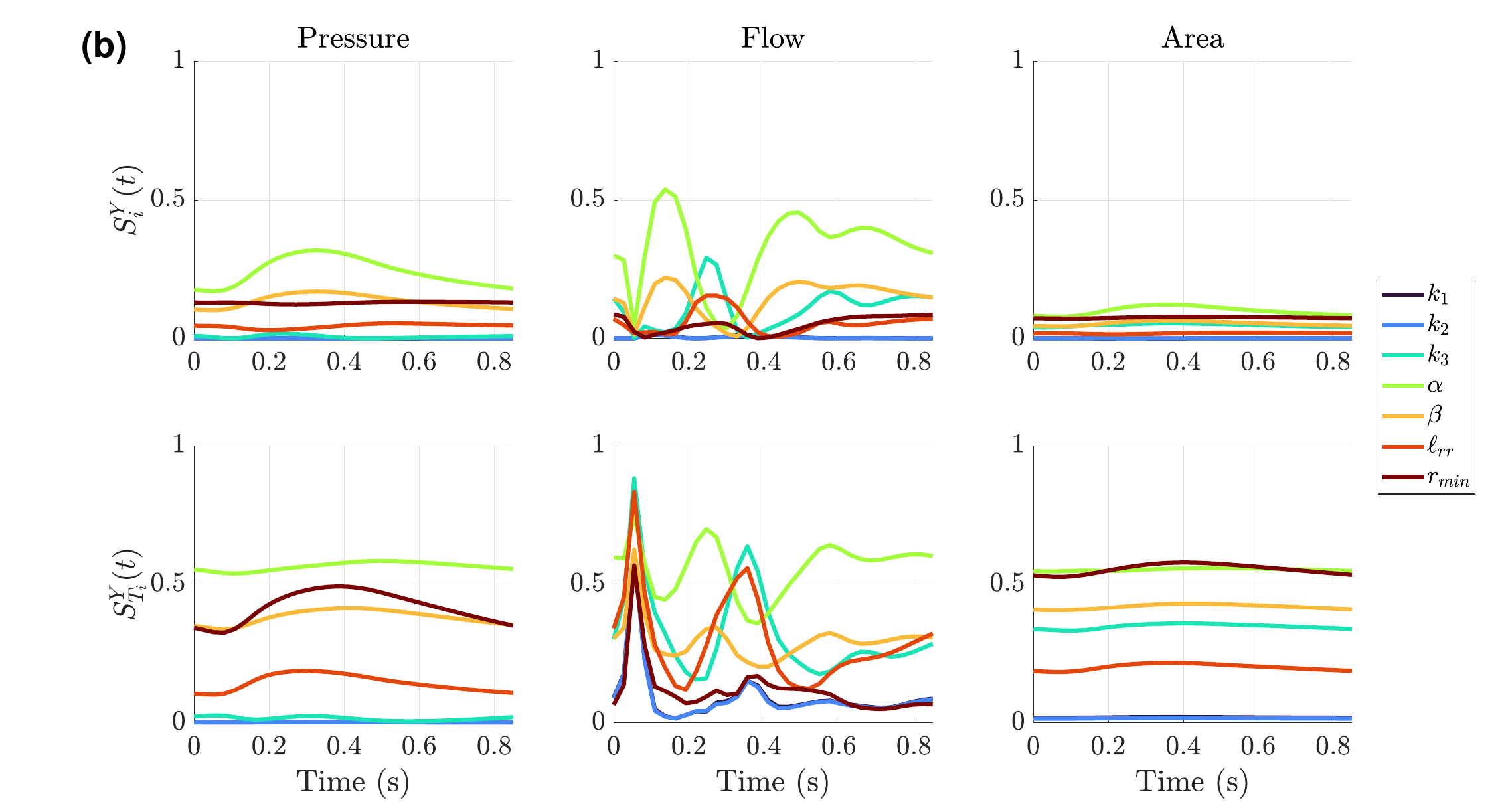}\vspace{-4mm}
        \includegraphics[width=0.9\linewidth]{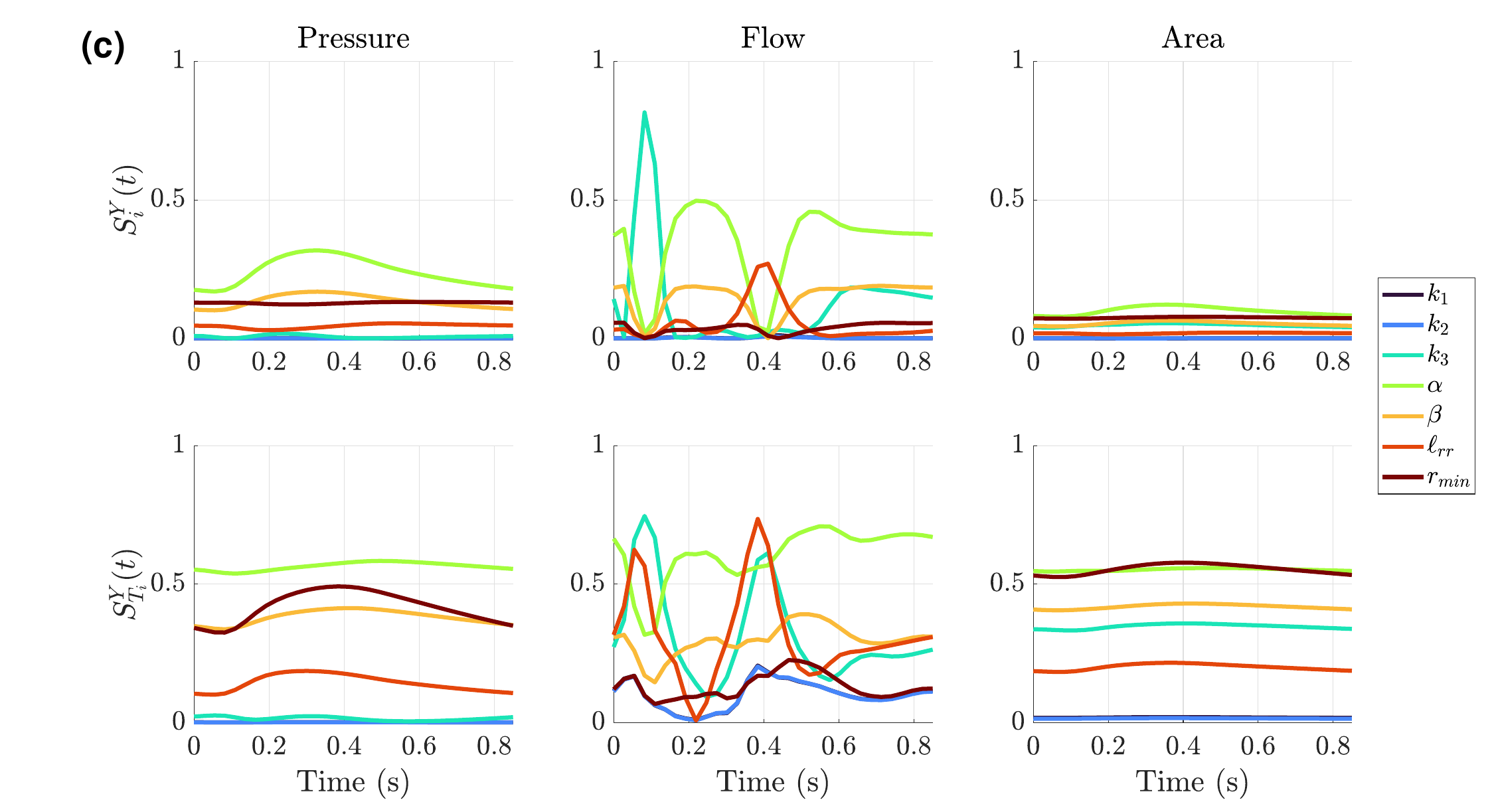}
    \end{subfigure}
    
    \caption{Point-wise Sobol' indices for the Structured tree boundary conditions calculated from principal component Sobol' metrics $S_i^{q}$ and $S_{T_i}^{q}$ via eq \eqref{eq:sob_time}. Results are presented for the (a) MPA, (b) LPA, and (c) RPA in the pulmonary tree.}
    \label{fig:Sobol_ST_time}
\end{figure}

We provide the point-wise estimates $S_i^{Y}(t)$ and $S_{T_i}^{Y}(t)$ in Figure \ref{fig:Sobol_ST_time}. For the pressure, it is clear that $\alpha$ is most influential, followed by $r_{min}$, $\beta$, and $\ell_{rr}$. The stiffness parameters $k_3$ shows some small changes in model sensitivity for the total order index. The shape of the sensitivity metrics tend to follow the typical time course of a pressure prediction from the model.

For the flow, we see several differences for the three branches. For the MPA (Figure \ref{fig:Sobol_ST_time}(a)), the stiffness $k_3$ dominates during the upstroke in systole, while $\alpha$ and $\beta$ become most influential following this initial increase in flow. The parameter $\ell_{rr}$ becomes most influential during the start of diastole, where $k_3$ and $r_{min}$ also become more influential. During diastole, the parameters $\alpha$ and $\beta$ become most influential. For the LPA flow (Figure \ref{fig:Sobol_ST_time}(b)), $S_i^{Y}(t)$ suggests that $\alpha$ is most influential for most of the cardiac cycle (except for a brief period where $k_3$ is largest). The parameters $\beta$, $k_3$, $r_{min}$, and $\ell_{rr}$ show larger $S_i^Y(t)$ values in diastole. The curves for $S_{T_i}^{Y}(t)$ in the LPA show an increase in all parameters, even $k_1$ and $k_2$ during the upstroke in flow. Here, $k_1$ and $k_2$ are still influential even through diastole, though they are smaller than the other parameters. Again we see $\alpha$ having largest influence on the flow on average, with $k_3$ and $\ell_{rr}$ increasing in influence during the start of diastole. Lastly, RPA flow results (Figure \ref{fig:Sobol_ST_time}(c)) show that $k_3$ has larger $S_{i}^{Y}(t)$ and $S_{T_i}^{Y}(t)$ during the upstroke in systole. For $S_{T_i}^{Y}(t)$ values, we see that $k_1$ and $k_2$ are minimally influential, with only some effects during the start of diastole. The parameters $\ell_{rr}$ appears largely influential during diastole as well.

Finally, the area sensitivity appears nearly identical for all three vessels. In general, $r_{min}$ and $\alpha$ are the most influential, with $\beta$ and $k_3$ being the next most influential. We see a relatively large difference in $S_{i}^{Y}(t)$ and $S_{T_i}^{Y}(t)$ for area (similar to results in Figure \ref{fig:Sobol_PCA}(b,d,f)) suggesting higher order interactions between parameters.

    

\subsection*{Profile-likelihood: Windkessel Model}
\begin{figure}[h!]
\centering
    \begin{subfigure}[b]{0.95\textwidth}
         \centering
        \includegraphics[width=0.9\linewidth]{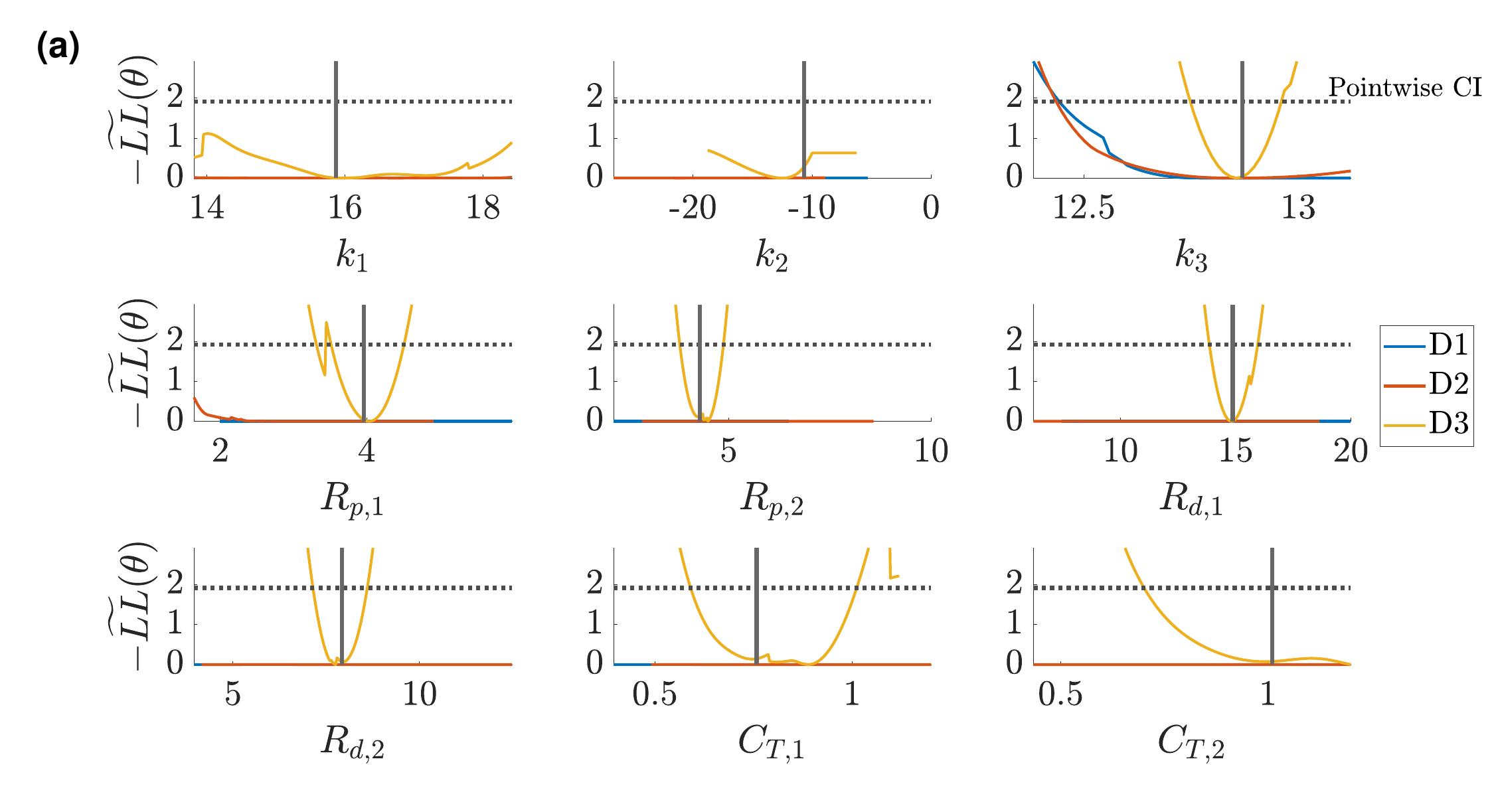}\\ \vspace{-4mm}
        \includegraphics[width=0.9\linewidth]{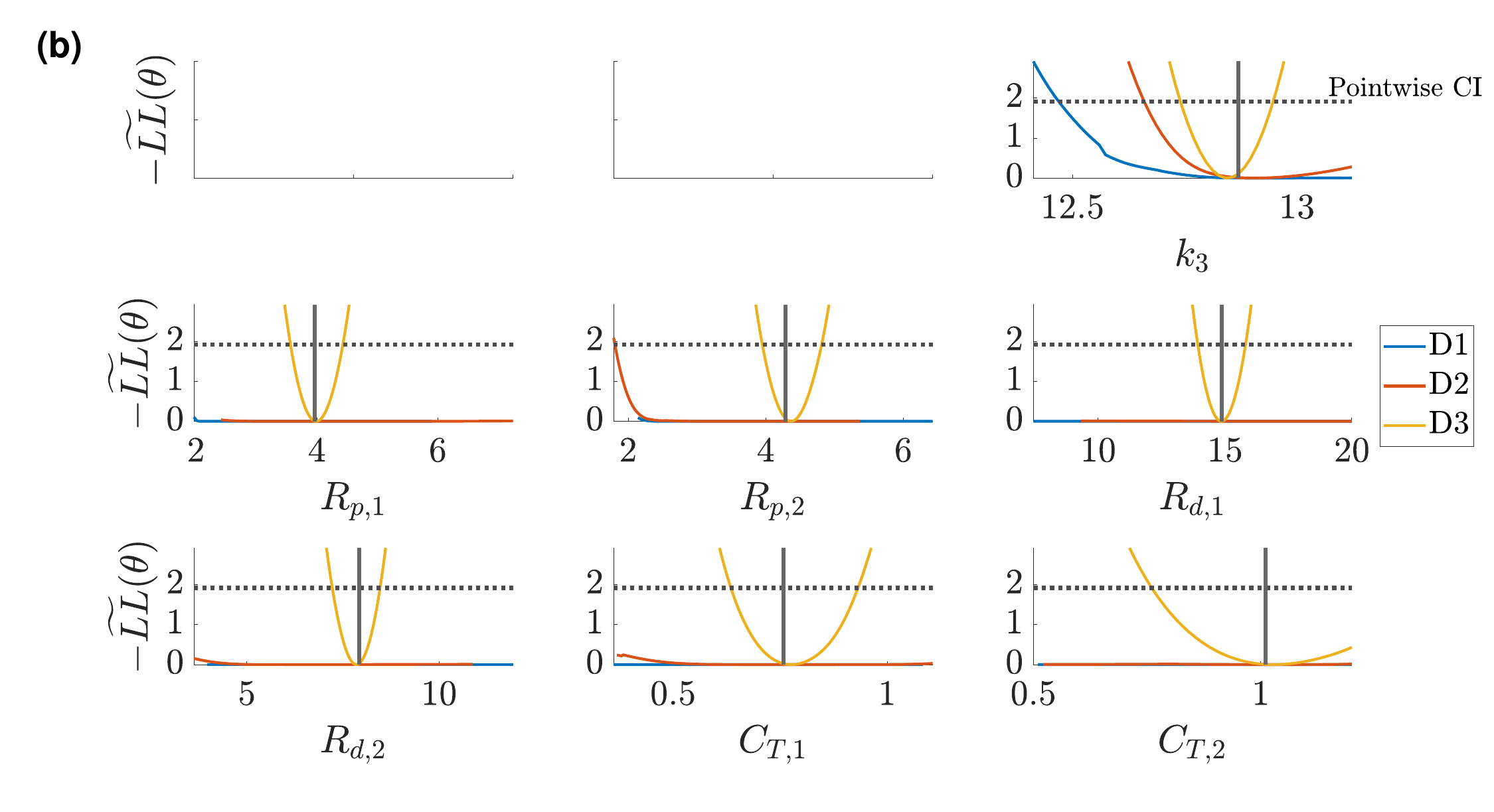}\\ \vspace{-4mm}
        \includegraphics[width=0.9\linewidth]{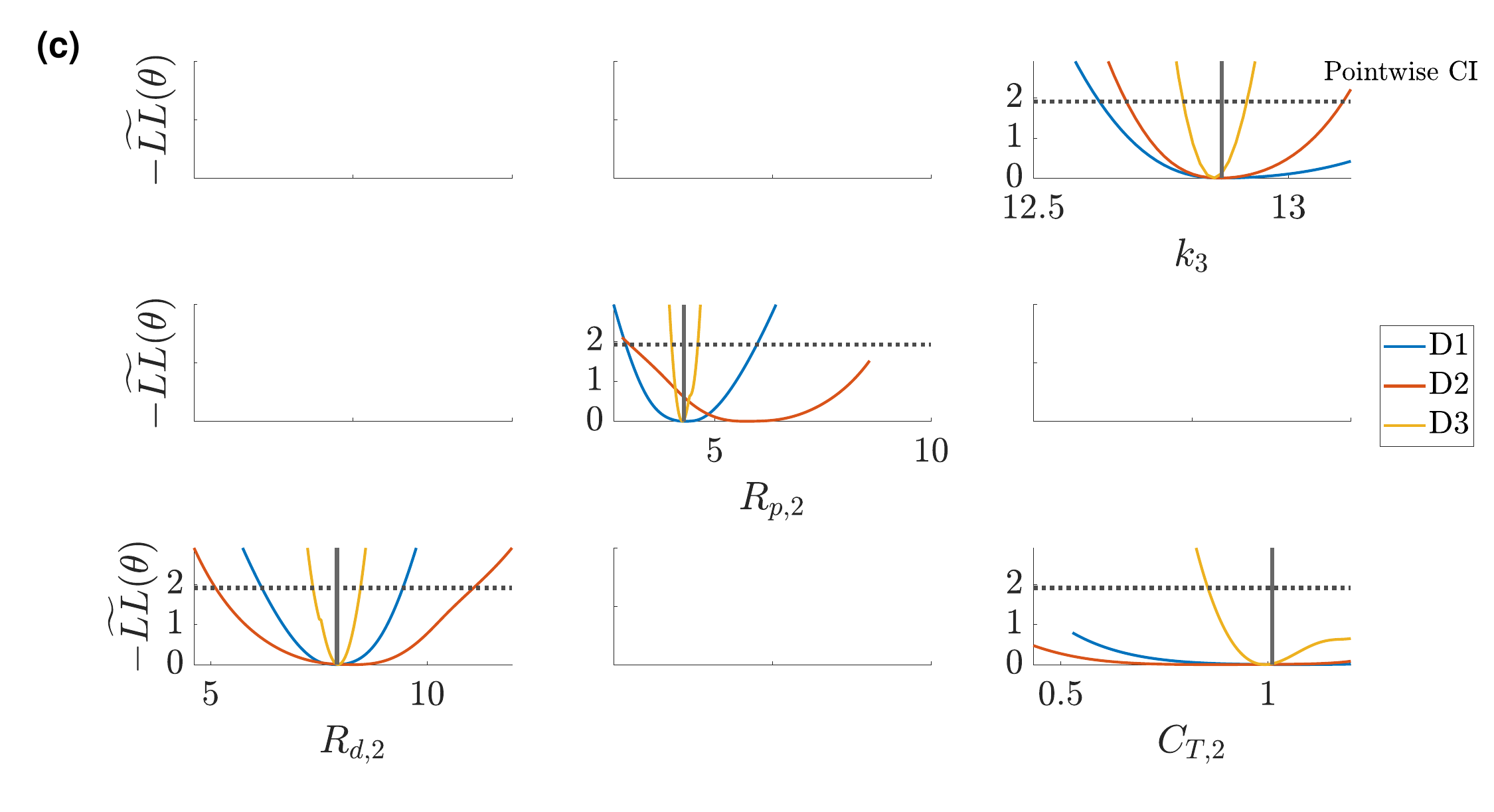}
    \end{subfigure}
    
    \caption{Profile-likelihood results using the PCA-PCE spectral surrogate. (a) Profile likelihood calculated using the three different experimental designs defined previously. The pointwise confidence intervals (defined in eq \eqref{eq:PL}) define whether parameters are considered identifiable. (b) A reduced parameter subset where $k_1$ and $k_2$ are not included in the profile-likelihood calculation. (c) A further reduced parameter set where $k_1$ and $k_2$ are fixed, as well as the LPA Windkessel parameters ($R_{p,1}$, $R_{d,1}$, and $C_{T,1}$). Blank plots represent parameters that are fixed.}
    \label{fig:PL_WK}
\end{figure}

We use the PCA-PCE spectral surrogate to calculate profile-likelihood confidence intervals for each design described previously. Figure \ref{fig:PL_WK} shows three sets of profile-likelihood results for one representative test dataset used for assessing model accuracy in Figure \ref{fig:testdata}. Results for the four other test data sets are provided in the Supplementary material \textbf{S1 File}. We first consider inferring all nine parameters (Figure \ref{fig:testdata}(a)) across the three different experimental designs. As expected, the parameters $k_1$ and $k_2$ have nearly flat profile-likelihoods for all three designs, although design $D3$ (MPA pressure and flows in the LPA and RPA) shows some increase in the negative log-likelihood. In general, all nine parameters are not identifiable for designs $D1$ and $D2$ which uses MPA pressure alone and MPA pressure with dynamic area in the LPA and RPA, respectively. The only exception is $k_3$, which has a single finite confidence bound. When using $D3$, parameters $k_3$, $R_{p,1}$, $R_{p,2}$, $R_{d,1}$, $R_{d,2}$, and $C_{T,1}$ all have finite confidence bounds, indicating they are all identifiable. The parameter $C_{T,2}$ is not practically identifiable due to its behavior for increasing values away from the minimum, but may have a finite confidence bound for a larger optimization range. 

We subsequently rerun our profile-likelihood analysis with $k_1$ and $k_2$ fixed, given their lack of identifiability and relatively small influence via sensitivity analysis. Results in Figure \ref{fig:testdata}(b) show similar trends to the full parameter set: only $k_3$ has one set of finite confidence bounds for $D1$ and $D2$. This time, $R_{p,2}$ also shows a finite confidence bound for $D2$. Again we see that, when using $D3$, nearly all the parameters have finite confidence bounds except for $C_{T,2}$. 

We hypothesize that this finding is related to interactions between the LPA and RPA boundary condition parameters, i.e., left and right boundary conditions may not be jointly identifiable. To test this, we ran an additional experiment where the parameters of the LPA ($R_{p,1}$, $R_{d,1}$, and $C_{T,1}$) were fixed. As shown in Figure \ref{fig:testdata}(c), this improved the identifiability of all four parameters. In particular, $k_3$, $R_{p,2}$ and $R_{d,2}$ are practically identifiable for all three designs ($R_{p,2}$ will be practically identifiable for $D2$ with a larger optimization range). In contrast, $C_{T,2}$ is still not practically identifiable. We performed an additional analysis (not shown) to further investigate whether fixing both compliance parameters would increase identifiability of the remaining parameters, but still found that only $D3$ provided full parameter identifiability.

\begin{figure}[h!]
\centering
    \begin{subfigure}[b]{0.95\textwidth}
         \centering
        \includegraphics[width=0.9\linewidth]{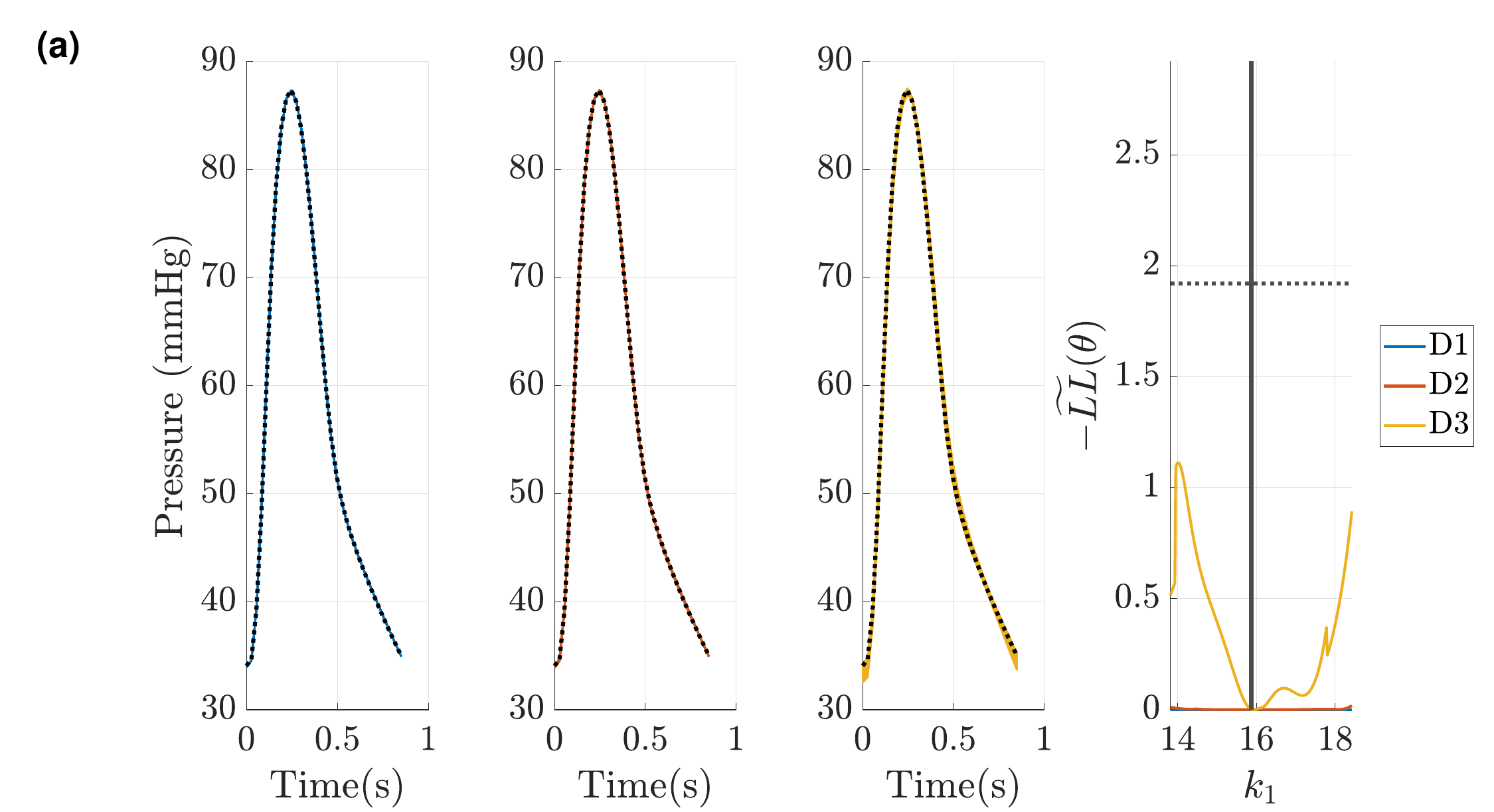}\vspace{-1mm}
        \includegraphics[width=0.9\linewidth]{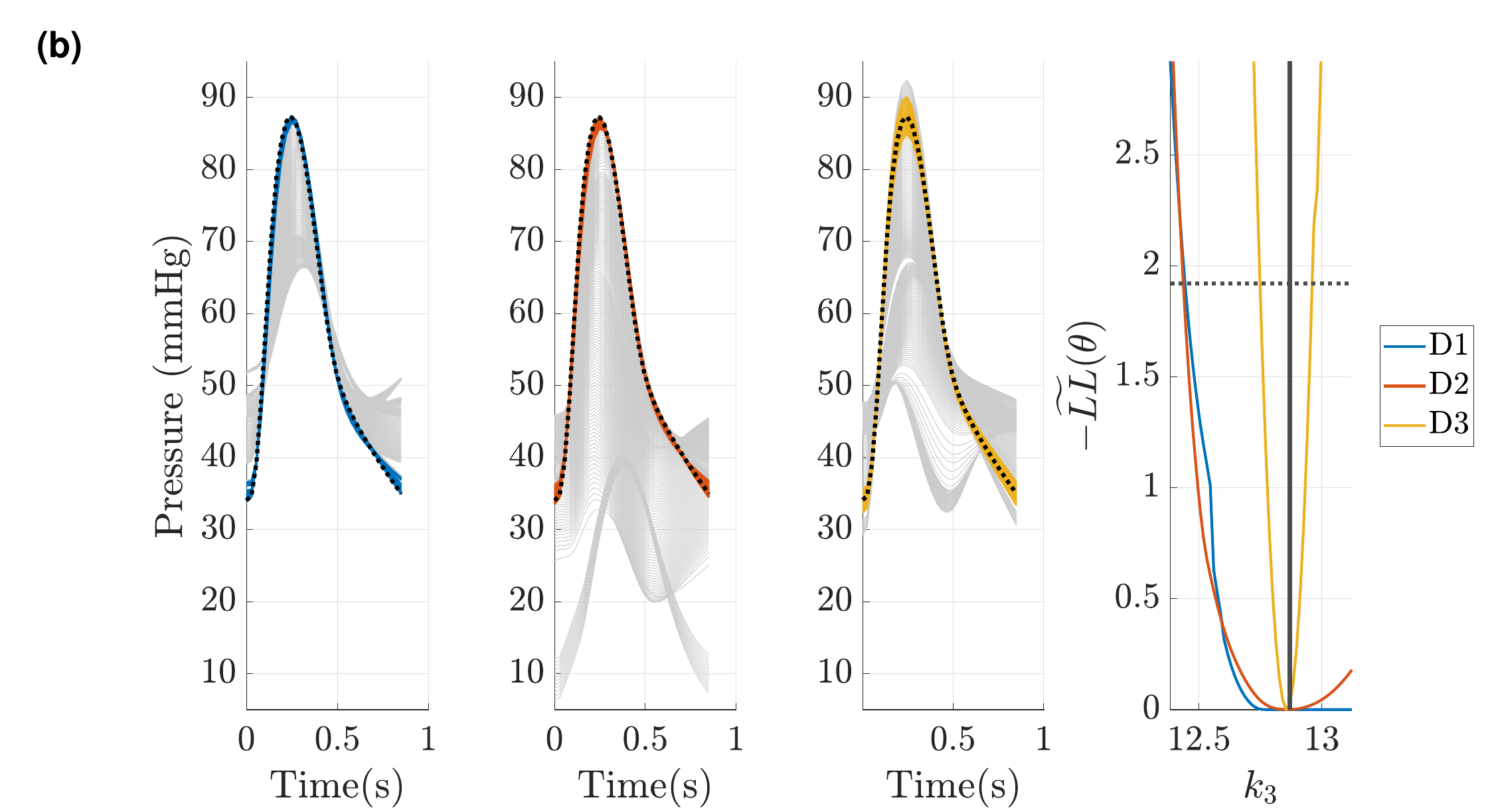}\vspace{-1mm}
        \includegraphics[width=0.9\linewidth]{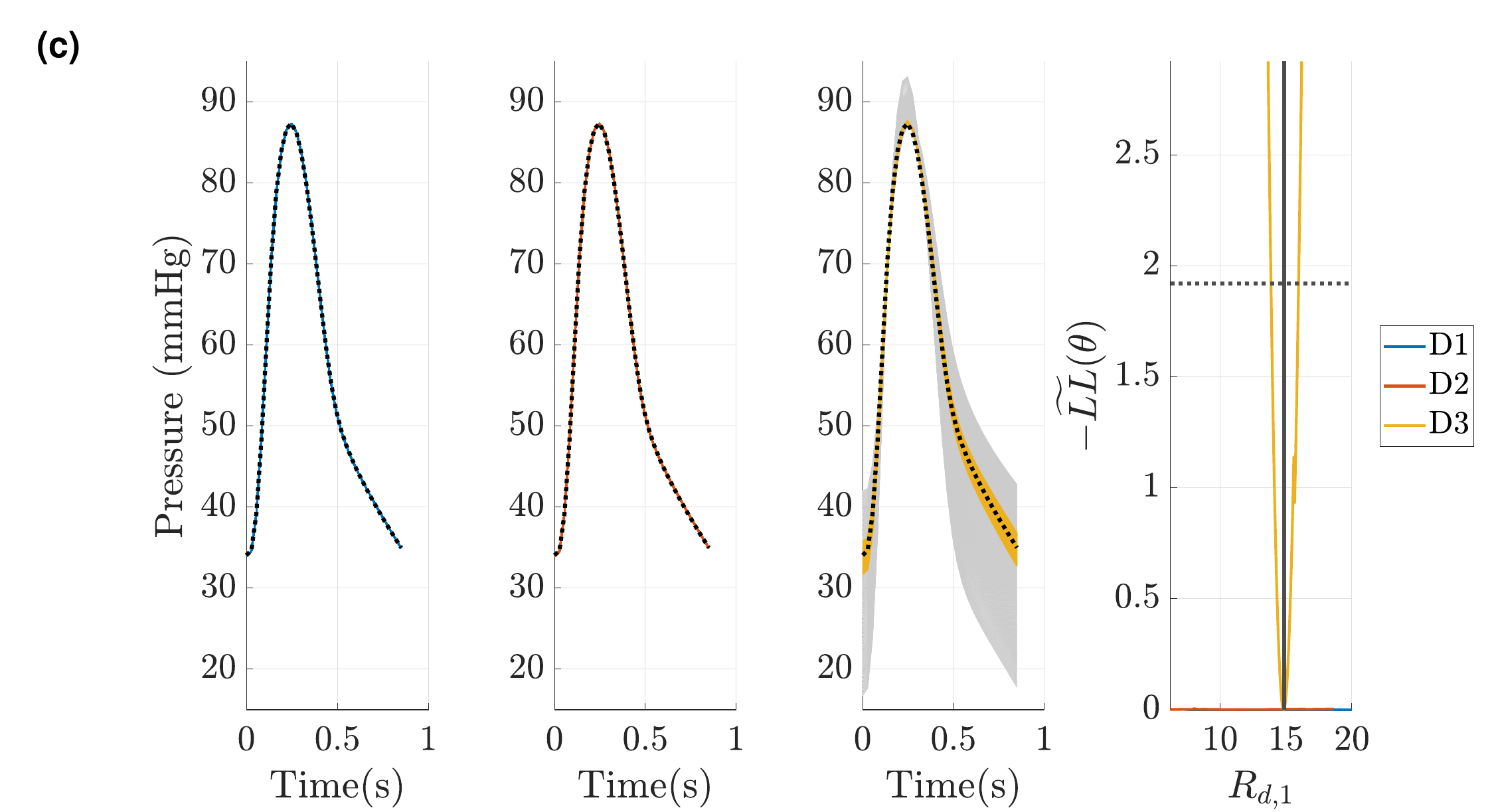}
    \end{subfigure}
    \caption{Model evaluations along the profile likelihood presented in Figure \ref{fig:PL_WK}(a) using designs $D1, D2$, and $D3$. Grey predictions represent all model parameters generated during the profile-likelihood, while color-coded predictions are model evaluations within the confidence interval threshold (shown as a dashed horizontal line in the rightmost subplot). The dashed black line represents the true signal used for calibration. (a) Evaluations along the $k_1$ profile likelihood. (b) Evaluations along the $k_3$ profile likelihood. (c) Evaluations along the $R_{d,1}$ profile likelihood.}
    \label{fig:PL_eval_WK}
\end{figure}

To emphasize how the profile-likelihood confidence intervals translate to signals in output space, Figure \ref{fig:PL_eval_WK} shows MPA pressure predictions from the inverse PCA transformed spectral surrogate using parameter values obtained along the profile likelihood in Figure \ref{fig:PL_WK}(a). Parameter $k_1$ (Figure \ref{fig:PL_eval_WK}(a)) is not identifiable, and surrogate predictions are indistinguishable along the profile-likelihood. In contrast, $k_3$ (Figure \ref{fig:PL_eval_WK}(b)) is only slightly practically non-identifiable (unbounded for larger $k_3$ values) for designs $D1$ and $D2$, and identifiable for $D3$. Note that for the latter, surrogate predictions vary greatly, suggesting stronger identifiability of the parameter. Finally, $R_{d,1}$ (Figure \ref{fig:PL_eval_WK}(c)) is practically non-identifiable for $D1$ and $D2$ (flat profile-likelihood), but is identifiable using $D3$. Model outputs support this claim, overlapping in $D1$ and $D2$, but varying in $D3$. The increase in identifiability (a larger variability in pressure predictions) is related to the change in experimental design, which incorporates different signals for calibration and profile-likelihood construction.

\subsection*{Profile-likelihood: Structured Tree Model}
\begin{figure}[h!]
\centering
    \begin{subfigure}[b]{0.95\textwidth}
         \centering
        \includegraphics[width=1\linewidth]{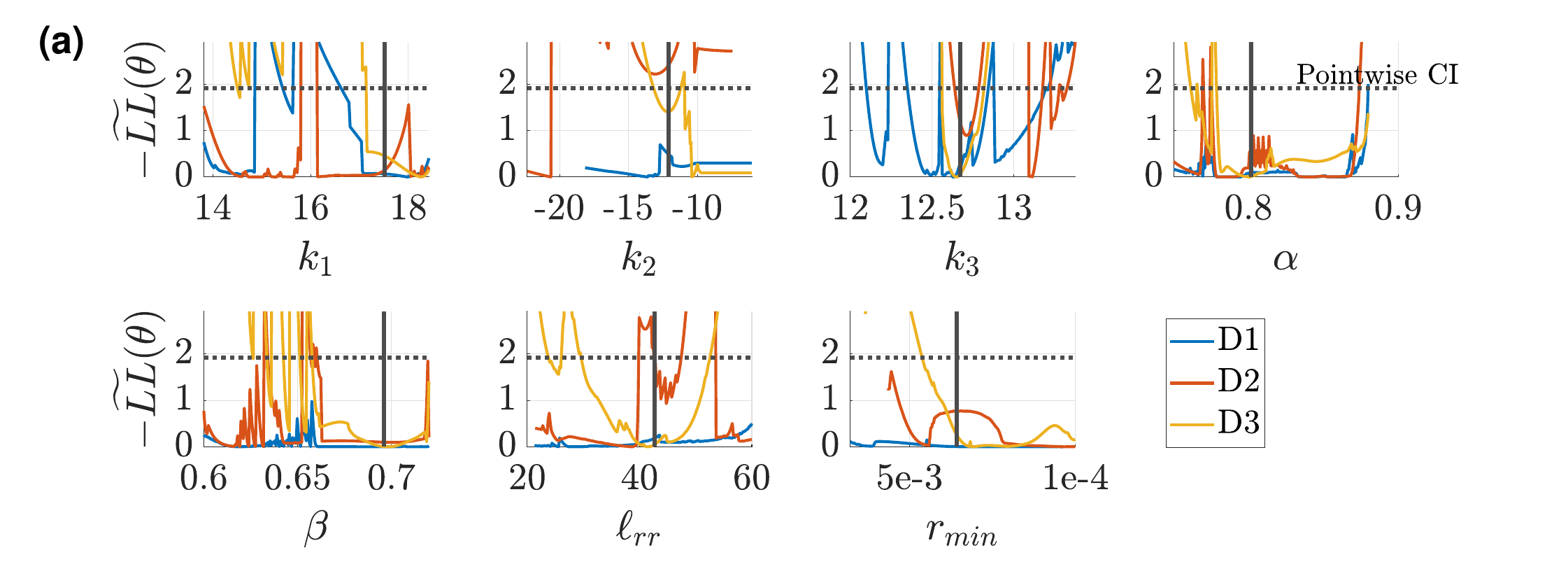}\\ 
        \includegraphics[width=1\linewidth]{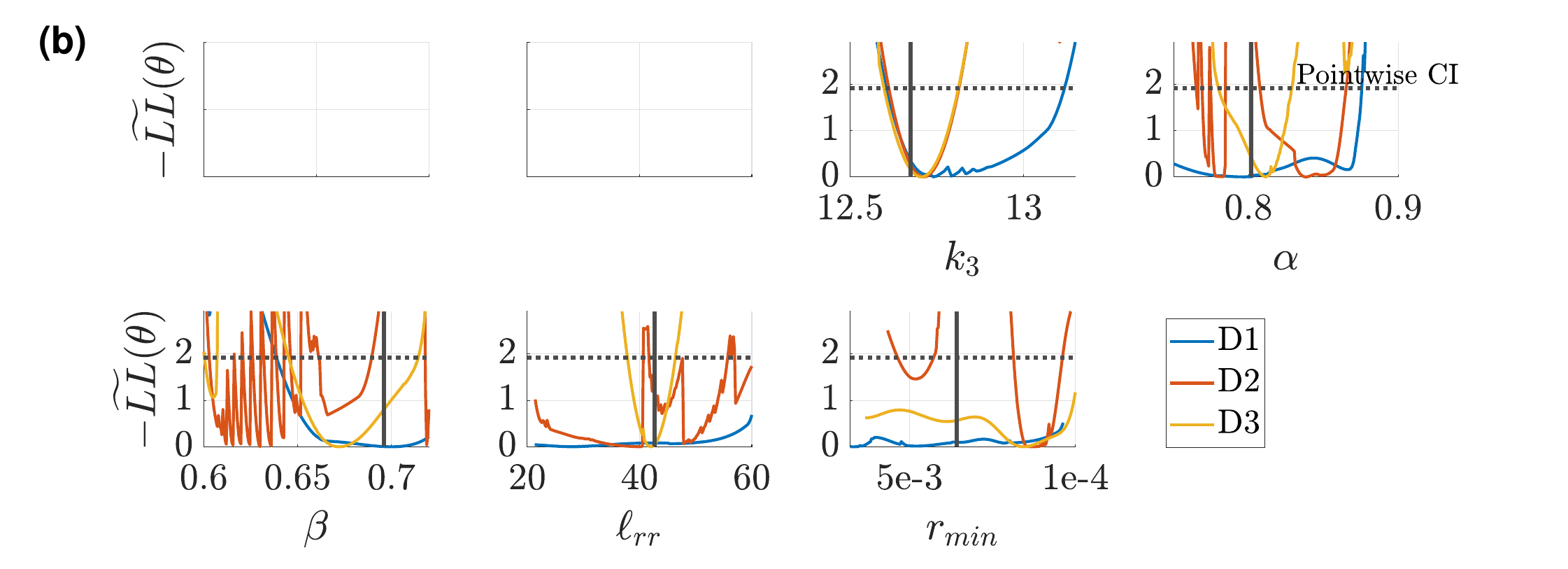}\\ 
        \includegraphics[width=1\linewidth]{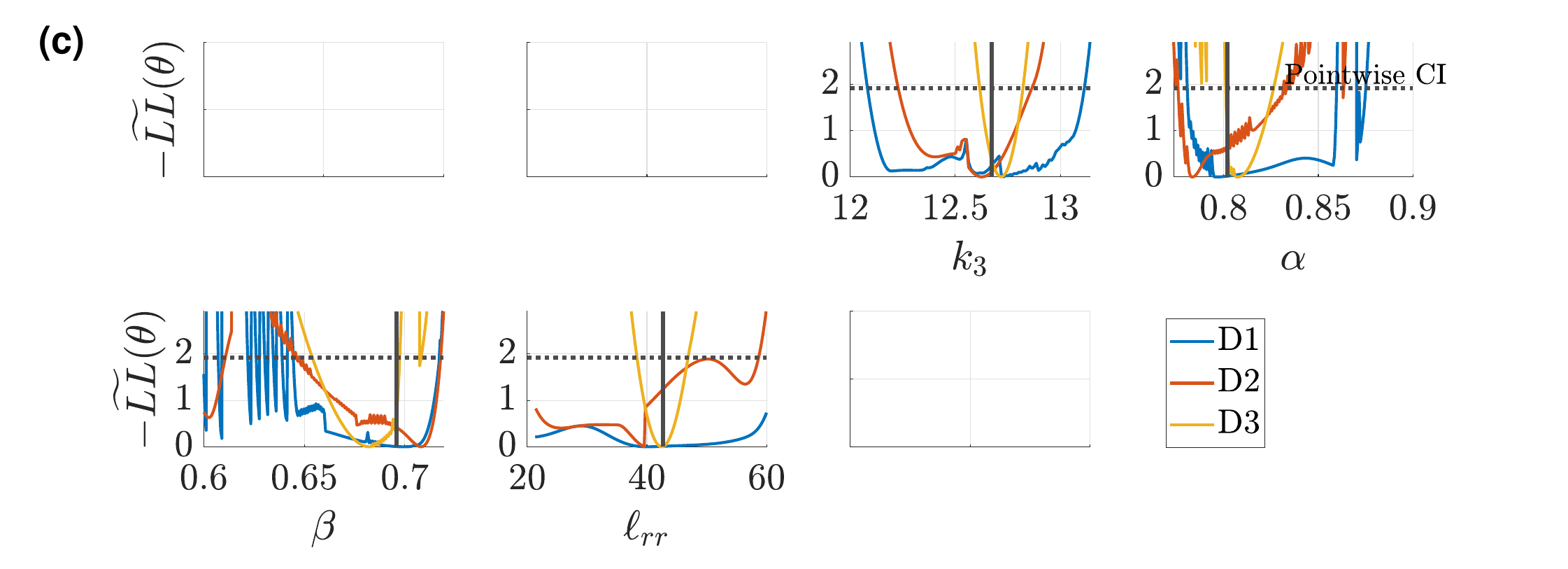}\\
        \includegraphics[width=1\linewidth]{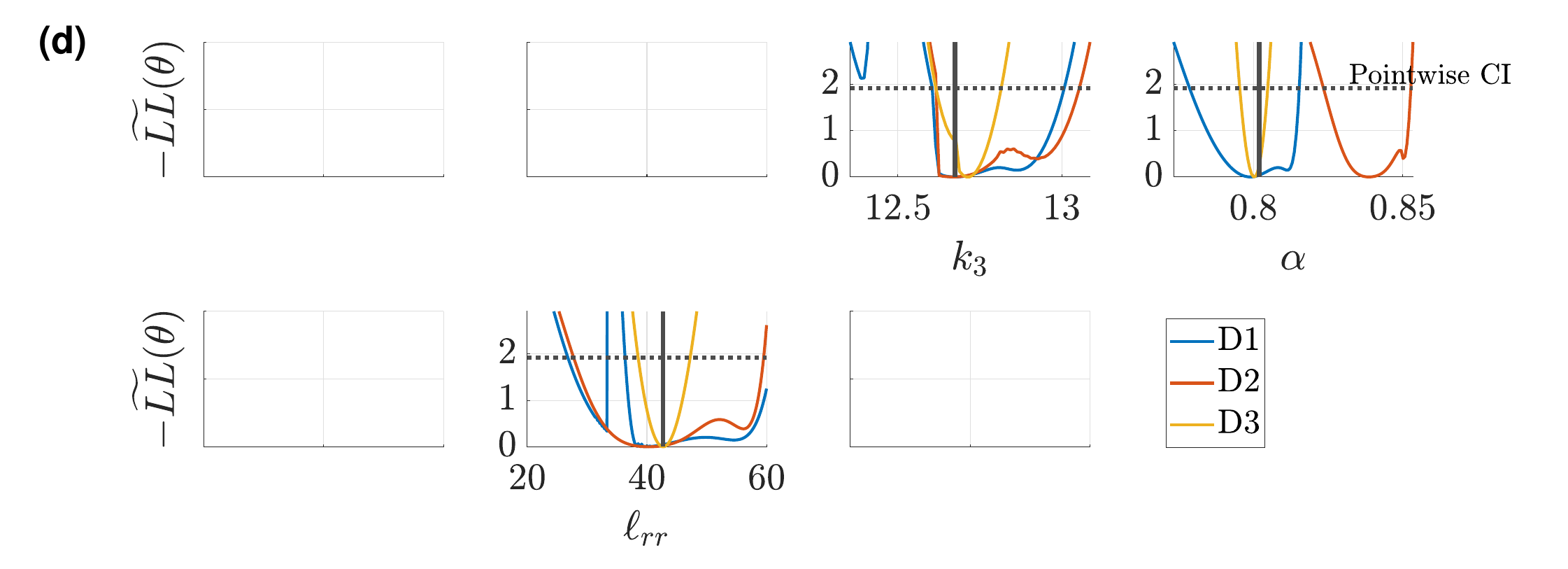}
    \end{subfigure}
    
    \caption{Profile-likelihood results using the PCA-PCE spectral surrogate. (a) Profile likelihood calculated using the three different experimental designs defined in previously. The pointwise confidence intervals (defined in eq \eqref{eq:PL}) define whether parameters are considered identifiable. (b) A reduced parameter subset where $k_1$ and $k_2$ are not included in the profile-likelihood calculation. (c) A further reduced parameter set where $\beta$ is fixed. (d) Further reduced parameter set with $r_{min}$ fixed. Blank plots represent parameters that are fixed.}
    \label{fig:PL_ST}
\end{figure}

We conduct a similar study using the structured tree boundary conditions. Figure \ref{fig:PL_ST}(a) shows the profile-likelihood results for all seven parameters using a representative test dataset. Results for the four other test data sets are provided in the Supplementary material \textbf{S1 File}. In contrast to the Windkessel boundary conditions, the likelihood values for the structured tree are more sporadic and include multiple local minima and/or possible discrepancies in the surrogate's ability to solve the inverse problem. It is difficult to discern whether any parameters satisfy the definition of identifiability, given the dynamic changes in $-\widetilde{LL}(\bm{\theta})$ values. We again consider fixing $k_1$ and $k_2$, as they are the least influential parameters and likely introduce identifiability issues. Results in Figure \ref{fig:PL_ST}(b) indicate that $k_3$ is now more identifiable, owing to the interactions between $k_1$ and $k_2$ during inference. The parameters $\alpha$, $\beta$, and $\ell_{rr}$ appear idenitifiable for $D3$, but are sporadic with multiple minima for other designs. The parameter $r_{min}$, though influential as shown in Figures \ref{fig:Sobol_PCA}(b,d,F) and \ref{fig:Sobol_ST_time}, appears difficult to identify. Hence, we fix $r_{min}$ and again conduct our profile-likelihood analysis. Figure \ref{fig:PL_ST}(c) shows improved identifiability for $\alpha$, $\beta$, and $\ell_{rr}$ for designs $D2$ and $D3$. As a final analysis, we consider the three dimensional parameter space with $k_3$, $\alpha$, and $\ell_{rr}$, as shown in Figure \ref{fig:PL_ST}(d). We fix $\beta$ based on its sporadic behavior. These results provide practical identifiability for all three parameters in $D3$, with minimal sporadic likelihood values. Interestingly, the parameter $\alpha$ is incorrectly inferred when using $D2$, and may reflect issues with inferring the full parameter set prior to calculating the profile-likelihood confidence intervals.

\begin{figure}[h!]
\centering
    \begin{subfigure}[b]{0.95\textwidth}
         \centering
        \includegraphics[width=0.9\linewidth]{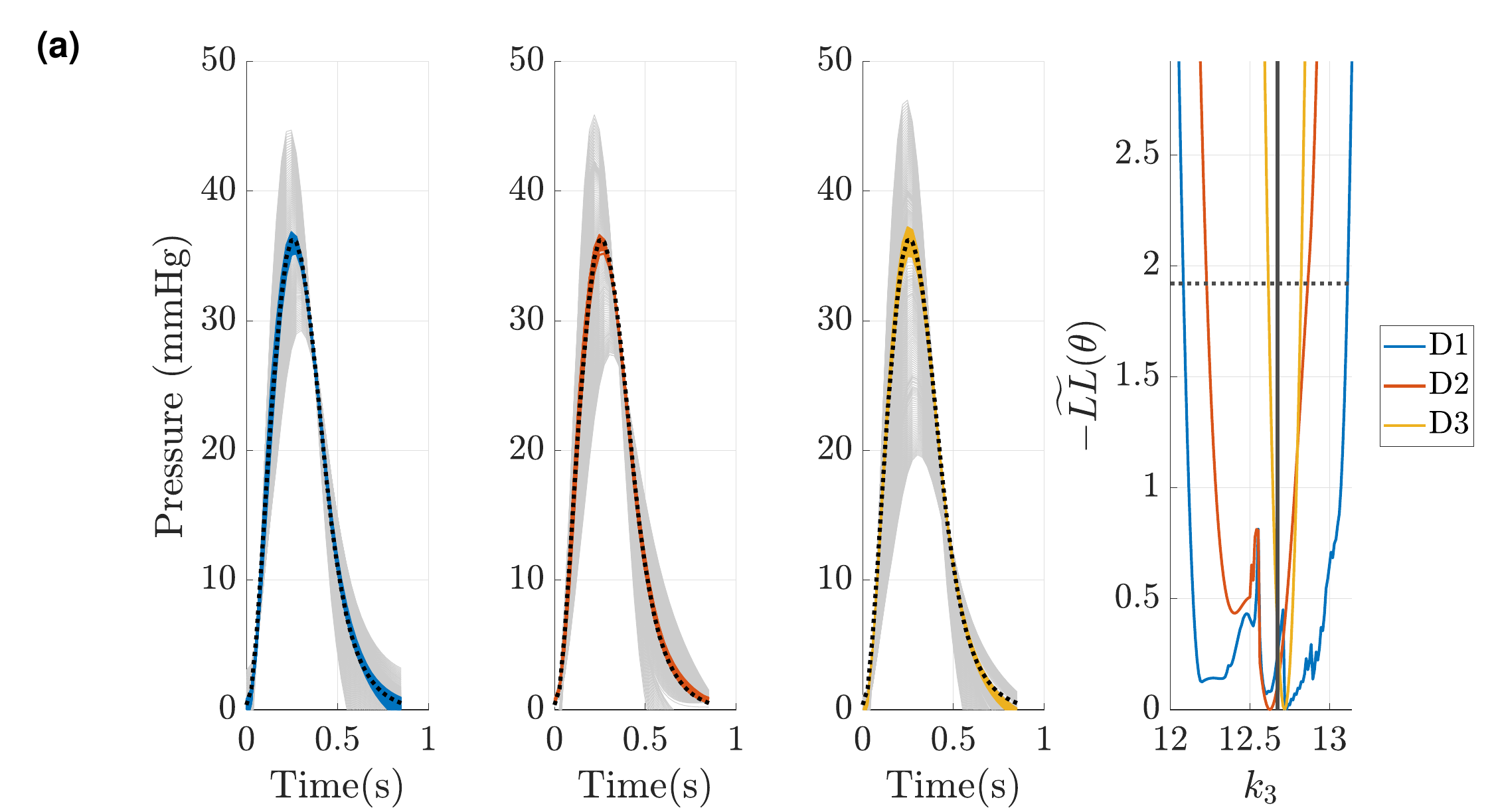}\vspace{-1mm}
        \includegraphics[width=0.9\linewidth]{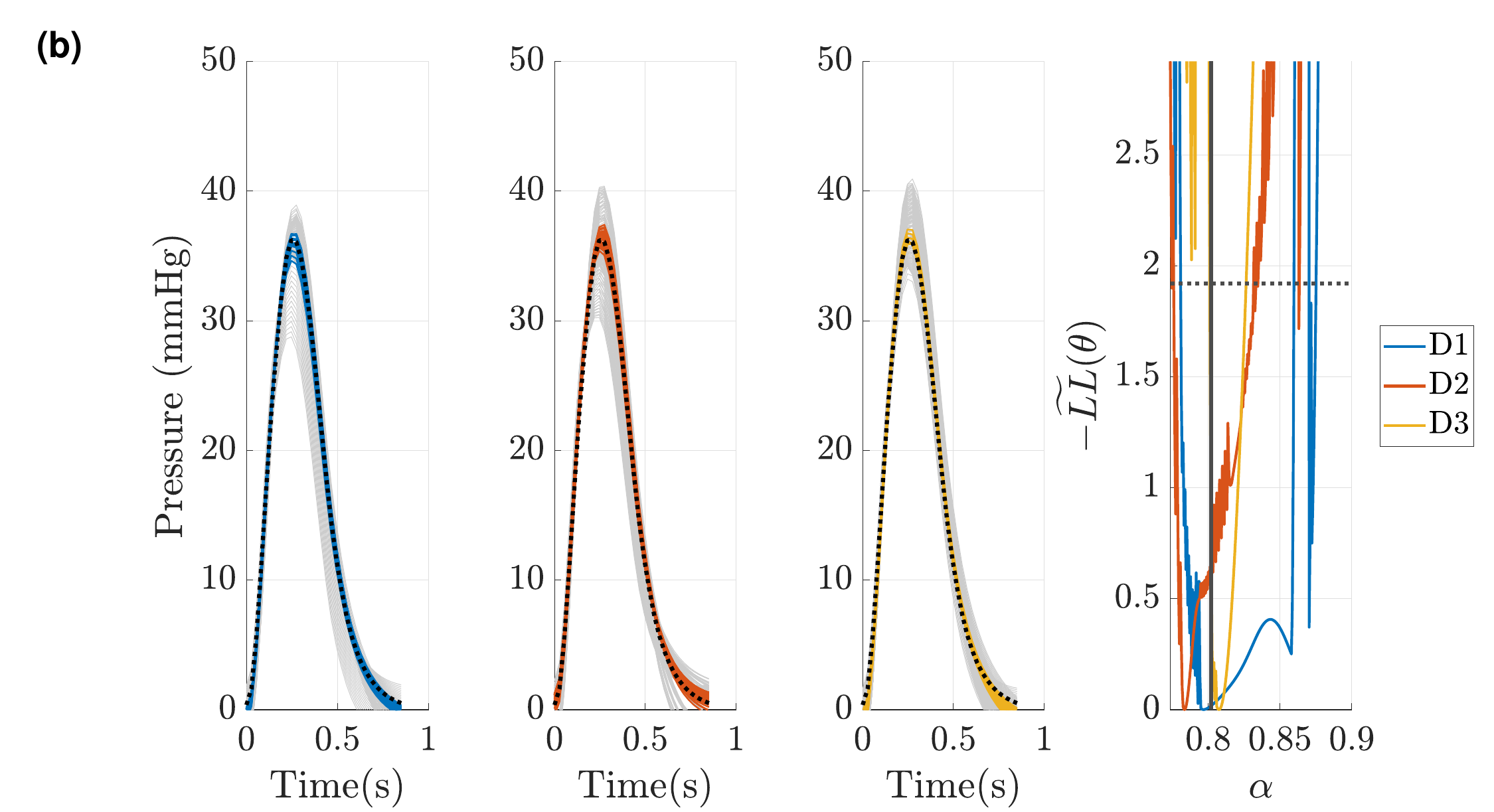}\vspace{-1mm}
        \includegraphics[width=0.9\linewidth]{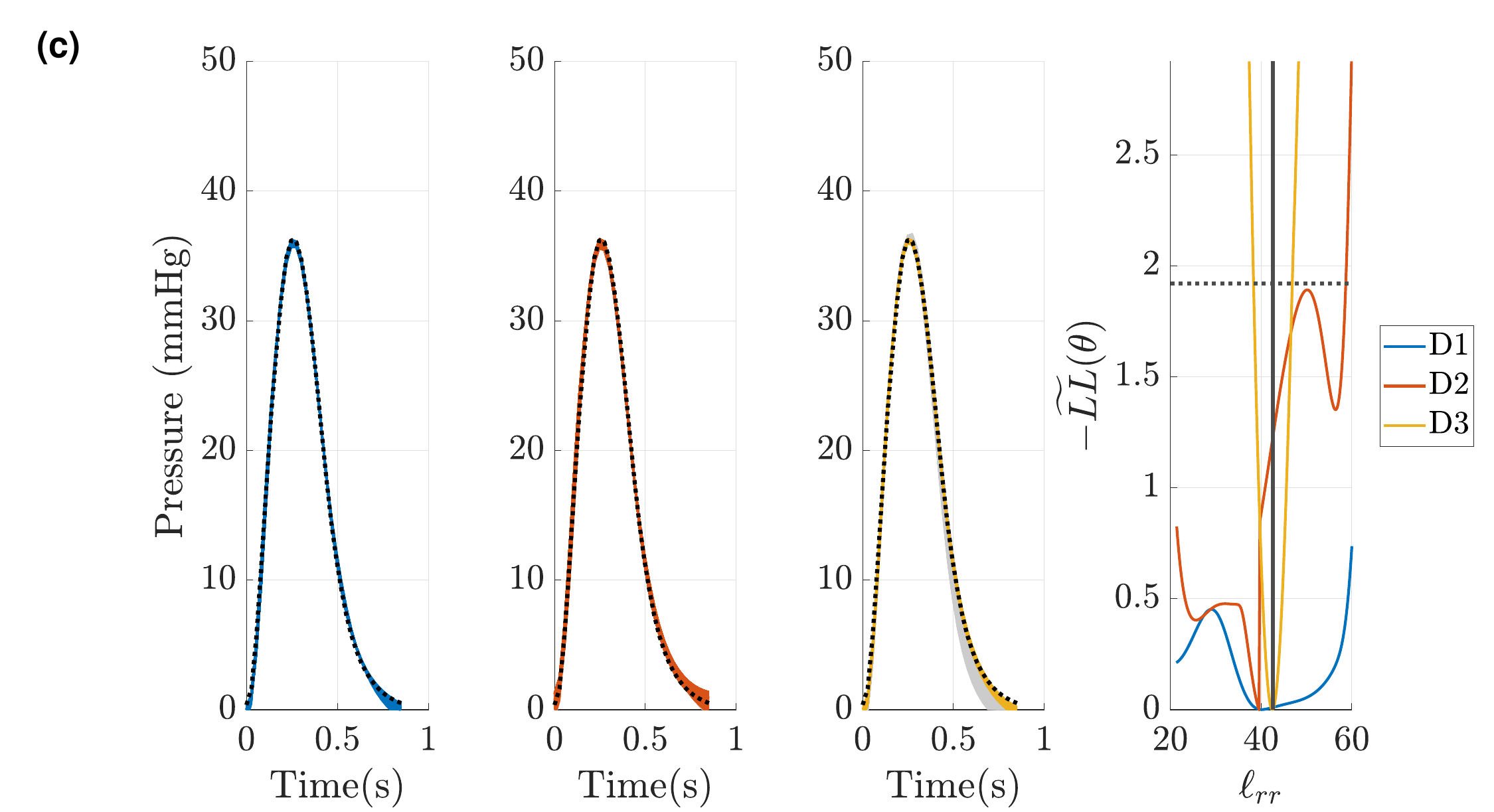}
    \end{subfigure}
    \caption{Model evaluations along the profile likelihood presented in Figure \ref{fig:PL_ST}(c) using designs $D1, D2$, and $D3$. Grey predictions represent all model parameters generated during the profile-likelihood, while color-coded predictions are model evaluations below the confidence interval threshold (shown as a dashed horizontal line in the rightmost subplot). The dashed black line represents the true signal used for calibration. (a) Evaluations along the $k_3$ profile likelihood. (b) Evaluations along the $\alpha$ profile likelihood. (c) Evaluations along the $\ell_{rr}$ profile likelihood.}
    \label{fig:PL_eval_ST}
\end{figure}

Similar to Figure \ref{fig:PL_eval_WK}, we provide output signals below and above the profile-likelihood confidence threshold in Figure \ref{fig:PL_eval_ST}. We use the four dimensional parameter space presented in Figure \ref{fig:PL_ST}(c), and provide output signals along the profile-likelihoods for $k_3$, $\alpha$, and $\ell_{rr}$. We see that both $k_3$ and $\alpha$ provide large variability in pressure signals above the confidence interval, whereas $\ell_{rr}$ provides only small deviations in pressure above the confidence threshold. The parameters $k_3$ and $\alpha$ are more influential on pressure than $\ell_{rr}$, hence they have larger deviations in output space.

\subsection*{Profile-likelihood: Simulator}
\begin{figure}[h!]
\centering
\hspace{-6cm}
    \begin{subfigure}[b]{\textwidth}
        \includegraphics[width=1.2\linewidth]{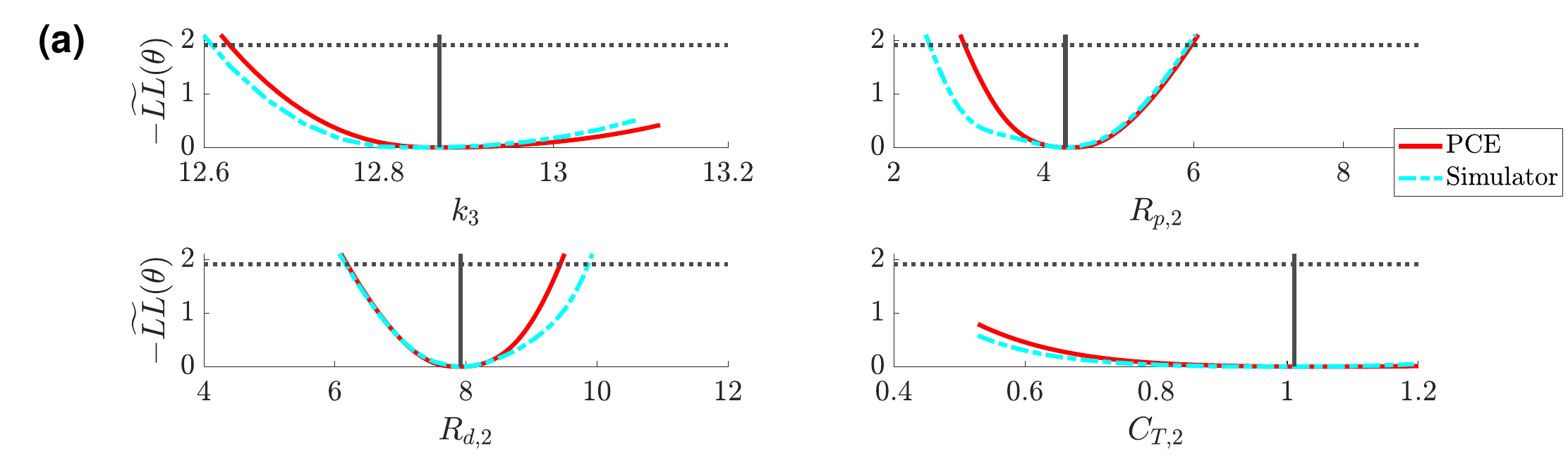}\vspace{-1mm}\\
        \includegraphics[width=1.2\linewidth]{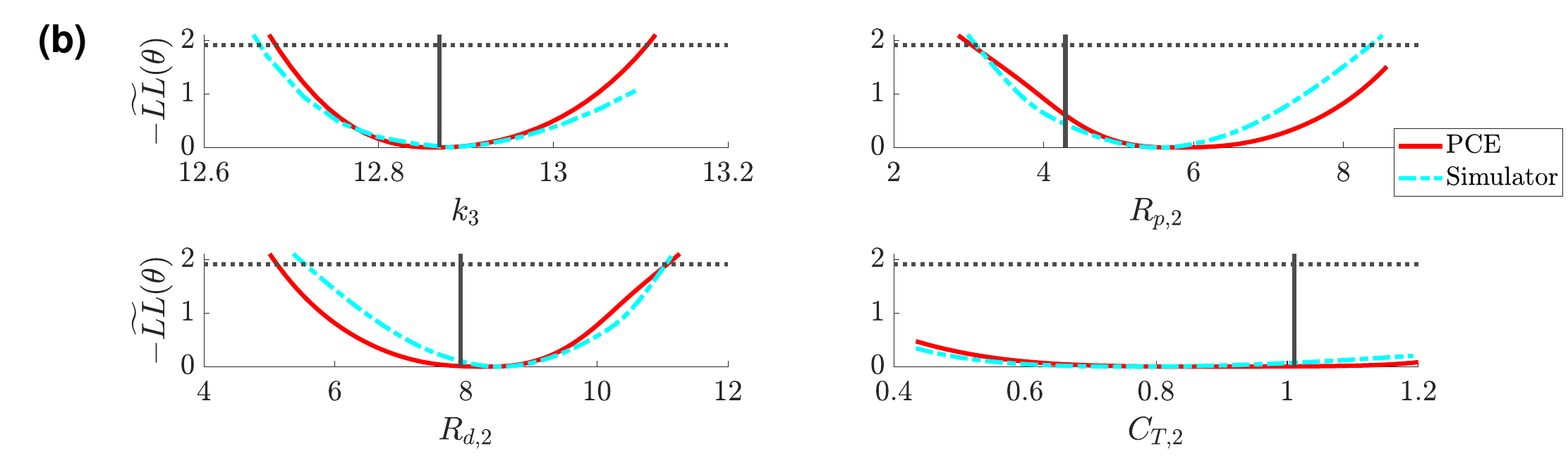}\vspace{-1mm} \\
        \includegraphics[width=1.2\linewidth]{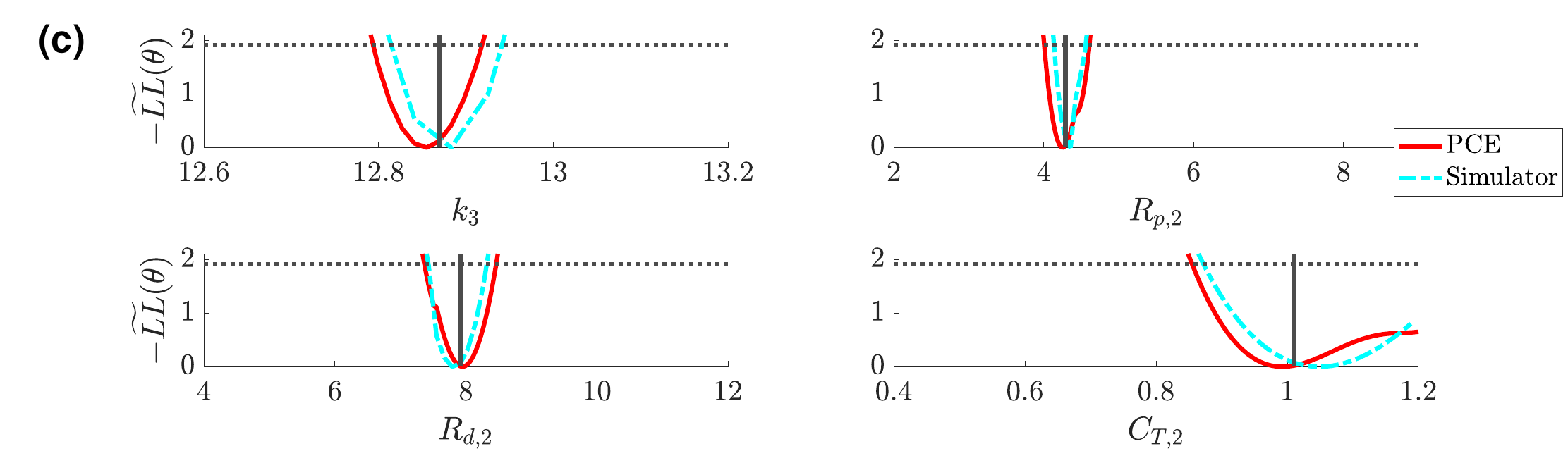}
    \end{subfigure}
    \caption{Comparison between the spectral surrogate (red) and simulator evlauted (cyan) profile-likelihood confidence intervals for the Windkessel boundary conditions using the free parameter set including $k_3$, $R_{p,2}$, $R_{d,2}$, and $C_{T,2}$. Results are shown for designs $D1$ (a), $D2$ (b), and $D3$
     (c).}
    \label{fig:PL_simulator_WK}
\end{figure}

\begin{figure}[h!]
\centering
    \begin{subfigure}[b]{\textwidth}
         \centering
        \includegraphics[width=1\linewidth]{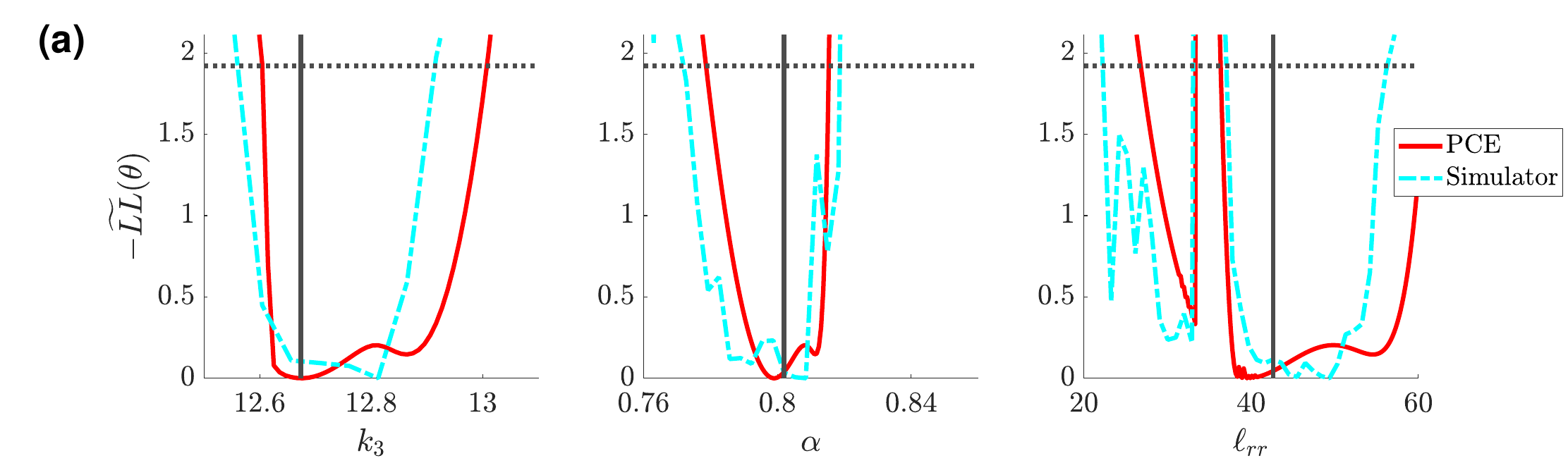}
        \includegraphics[width=1\linewidth]{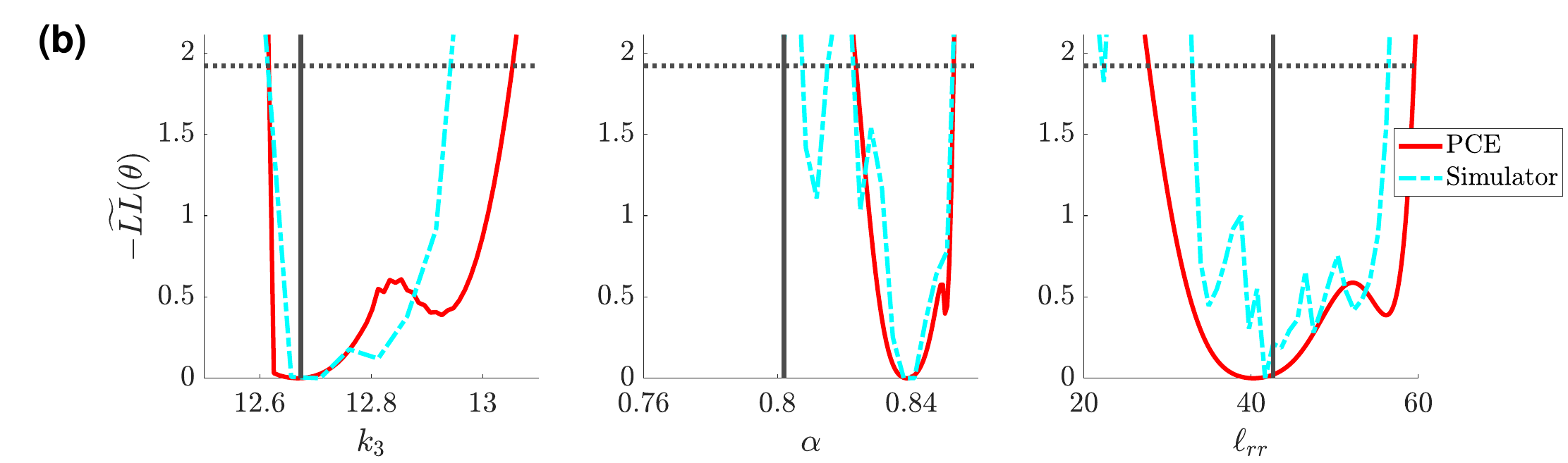}
        \includegraphics[width=1\linewidth]{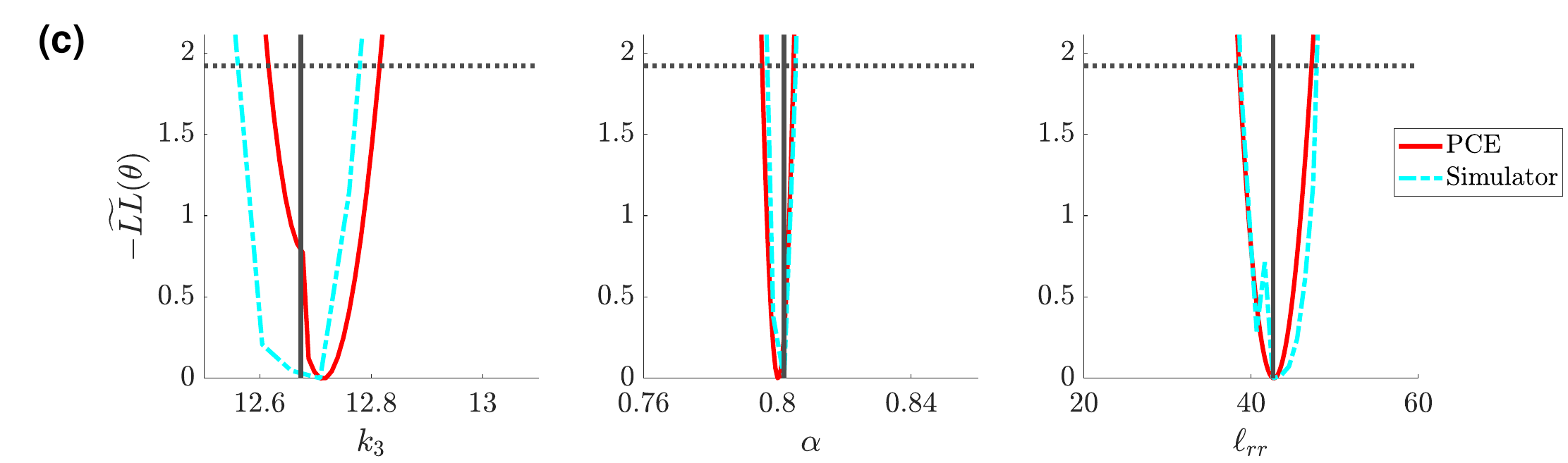}
    \end{subfigure}
    \caption{Comparison between the spectral surrogate (red) and simulator evaluated (cyan) profile-likelihood confidence intervals for the structured tree boundary conditions using the free parameter set including $k_3$, $\alpha$, and $\ell{rr}$. Results are shown for designs $D1$ (a), $D2$ (b), and $D3$
     (c).}
    \label{fig:PL_simulator_ST}
\end{figure}

The full PDE simulator is too expensive to use for profile-likelihood analyses, hence the use of a spectral surrogate. However, we can use the resulting profile-likelihood parameter estimates to assess whether the values of $-\widetilde{LL}(\bm{\theta})$ are comparable with the true simulator. Figures \ref{fig:PL_simulator_WK} and \ref{fig:PL_simulator_ST} compare the PCA-PCE spectral surrogate values of the profile-likelihood compared with the PDE simulator derived likelihood for the Windkessel and structured tree boundary conditions, respectively. Each simulator curve is generated by passing the profile-likelihood parameters to the simulator and then calculating the same weighted likelihood function as used in the profile-likelihood calculation. For the Windkessel model, we investigate the profile-likelihood parameter values for the parameter subset including $k_3,R_{p,2},R_{d,2},$ and $C_{T,2}$ (similar to Figure \ref{fig:PL_WK}(c)). For the structured tree model, we investigate the profile-likelihood parameter values for the parameter subset including $k_3,\alpha,$ and $\ell_{rr}$ (similar to Figure \ref{fig:PL_ST}(d)). 

Though the confidence interval shapes do not match exactly, we observe that the conclusions about identifiability agree between the surrogate framework and the true simulator. Design $D3$ is the most informative design for both models, with $D2$ providing some additional information for parameter identification over $D1$ but not necessarily guaranteeing that parameters are identifiable. As expected, the Windkessel results appear more consistent between the simulator and emulator (which had a smaller emulation error), whereas the simulator results for the structured tree model show how the emulator and simulator outputs have non-smooth likelihood values. However, the results in Figures \ref{fig:PL_simulator_WK} and \ref{fig:PL_simulator_ST} show that the confidence bounds are similar when using the emulator and simulator.

\section*{Discussion}
We use spectral surrogates to determine parameter influence via sensitivity analysis and identifiability by constructing profile-likelihood confidence intervals. While multiple studies have used spectral surrogates in the context of pulse-wave propagation PDE models, none have considered extending these surrogates for formal identifiability analyses. In addition, we present a relatively simple procedure for handling vectorized outputs by reducing the output dimensionality using PCA. We use this framework to assess the sensitivity of different model outputs and test different experimental designs for parameter identifiability. This approach, though applied to a hemodynamics model, is applicable to broader classes of expensive PDE models, which have yet to formally adopt identifiability methods such as the profile-likelihood. Moreover, our study examines the two most common boundary conditions for hemodynamics models, providing insight for future studies looking to build subject-specific pulmonary models via parameter inference.

\subsection*{Spectral surrogates}
One of our major goals in this article was to illustrate how spectral surrogates and PCEs can be leveraged beyond global sensitivity analysis, requiring no additional simulator evaluations to address parameter identifiability.  Here we adopted a similar approach to Nagel et al. \cite{Nagel2020}, which combined PCEs with dimension reduction in the output space via PCA in the context of urban drainage simulation. A similar approach was considered in the study by Paun et al. \cite{Paun2025}, which compared forward emulation and inverse problem accuracy between PCEs and Gaussian processes for a similar PDE hemodynamics model. Paun et al. also compared full dimensional output emulation (i.e., for full time series) to a PCA reduced output, showing similar accuracy between full and PCA reduced outputs. 

The PCE surrogate provides similar accuracy across the five different outputs as show in Figure \ref{fig:testdata} for the Windkessel boundary conditions. There is a larger emulation error for the structured tree model, especially for the pressure output, with outliers on the order of 4\% to 15\%. This is attributed in part to the relatively larger variance achieved by the structured tree model when conducting sampling, as documented previously \cite{Paun2025}. The variances attributed to each principal component, shown in Table \ref{tab:Variance}, show that five principal components typically capture 99\% of the variance. The exceptions include flow in the RPA (vessel 3) for the Windkessel boundary conditions, as well as the two daughter flows in the structured tree model. However, the emulation accuracy for the flows across both models are relatively well captured, suggesting that even less than 99\% of the variance is necessary for building a sufficiently accurate spectral surrogate.

\subsection*{Sensitivity analysis: Windkessel model}
Variance based sensitivity analysis is considered the gold standard for global sensitivity analysis \cite{Eck2016,Nagel2020}, with numerous hemodynamic modeling studies employing Sobol' indices for uncertainty quantification \cite{Colebank2024,Donders2015,Eck2017,Huberts2014,Melis2017}. The use of dimension reduction to quantify the sensitivity of entire time series signals has been employed previously \cite{Nagel2020}, but, to the author's knowledge, has yet to be used for hemodynamics modeling. The study by Huberts et al. \cite{Huberts2014} used PCEs to identify the sensitivity of a 1D systemic hemodynamics model to vascular stiffness and Windkessel parameters among other parameters. They found that Windkessel resistances were typically influential for both mean flow and systolic radial artery pressure. A similar study by Donders et al. \cite{Donders2015} combined Morris screening (another sensitivity method) with Sobol' indices to quantify the impacts of model parameters on mean brachial flow and distal arterial systolic pressure. They found again that Windkessel resistance was influential, whereas compliance was only influential in certain sections of the circulation. The study by Melis et al. \cite{Melis2017} used Gaussian processes to emulate systolic and diastolic pressure in a systemic vascular network model, and again found that Sobol' indices were largest for resistance and, in some sections of the systemic vasculature, stiffness. These studies focused on a single quantity of interest for each sensitivity conclusion.

These previous findings are consistent with our results in Figure \ref{fig:Sobol_PCA}(a,c,e) and \ref{fig:Sobol_WK_time}, which show that compliances are on average the least influential parameters after the exponential stiffness terms $k_1$ and $k_2$. Similar findings came from the prior study by Colebank et al. \cite{Colebank2019} which found again that parameter related to compliance were the least influential on pulmonary artery pressure conditions. This is consistent with our pressure sensitivities in Figure \ref{fig:Sobol_WK_time}, while we also identify an increased role for the compliance parameters in the left and right pulmonary flows, as shown in Figure \ref{fig:Sobol_WK_time}(b-c). This implies that flow distribution is based in part on compliance values. We also see a dominance of stiffness $k_3$ in influence on area predictions in all three branches. This is intuitive, as stiffness plays a role in the relative area change and area magnitude given by the pressure-area relationship in eq \eqref{eq:state}. The time-series sensitivities in Figure \ref{fig:Sobol_WK_time} show that the model sensitivity fluctuates with the cardiac cycle. Similar to findings by Colebank et al. \cite{Colebank2019}, the results show that pressure sensitivity to distal resistance decreases during systole and then increases in diastole, while stiffness is more influential during systole. 
The flow sensitivities follow unique trajectories in all three branches, with the LPA and RPA flow sensitivities suggesting that resistance and compliance elements corresponding to the left and right side are most influential in the respective arteries. We note that the MPA flow is a boundary condition, though the MPA sensitivity is calculated at the midpoint whether there is some flow fluctuations. 

\subsection*{Sensitivity analysis: Structured tree model}
To date, relatively few studies have conducted a sensitivity analysis on models using the structured tree boundary conditions. The study by Perdikaris et al. \cite{Perdikaris2015} conducted a pseudo-sensitivity analysis with respect to the minimum radius, finding that large values of $r_{min}$ had drastic effects on pressure in the basilar and radial artery but relatively small effects on flow magnitude. Olufsen et al. \cite{Olufsen2012} conducted a similar pseudo-sensitivity analysis on both systemic and pulmonary circulation models using the structured tree, and saw that structured tree parameters linked to $\alpha$ and $\beta$ caused significant changes in MPA pressure shape and magnitude while stiffness had a large effect on MPA pulse-pressure. The study by Taylor-Lapole et al. \cite{TaylorLapole2025} used Morris screening on 1D hemodynamics models for subjects with congenital heart defects, and found that the $k_3$ was most influential on a combination of pressure and flow outputs, with $\alpha$, $\beta$, and $\ell_{rr}$ showing slightly smaller influence. The studies by Paun et al. \cite{Paun2025} and Colebank and Chesler \cite{Colebank2024} use Sobol' indices to rank parameters in arterial and arterio-venous structured tree models. They also find that $\alpha$, $\beta$, $\ell_{rr}$ and $r_{min}$ are most influential on pressure and flow, but find mixed results for the sensitivity to stiffness. 

Our findings in Figure \ref{fig:Sobol_PCA}(b,d,f) shows that  $\alpha$, $\beta$, $\ell_{rr}$ and $r_{min}$ are most influential on the first principal component of pressure, with $k_3$ being more infleutial on higher components. The flow and area outputs are more influential to stiffness, and thus supports the previous finding by Taylor-Lapole \cite{TaylorLapole2025} which found stiffness most influential for a combination of mostly flow outputs and some pressure outputs. Our time-dependent Sobol' values in Figure \ref{fig:Sobol_ST_time} show again that sensitivity magnitudes vary with the dynamics of the cardiac cycle. The $S_i$ and $S_{T_i}$ values for the pressure and area do not qualitatively vary much with the cardiac cycle. In contrast, the flow sensitivity clearly fluctuates with cardiac cycle across the MPA, LPA, and RPA. We focus attention on the LPA and RPA given the use of data as an MPA flow boundary condition. For the LPA (Figure \ref{fig:Sobol_ST_time}(b)), we see that $\alpha$ and $\beta$ are most influential via $S_i$ during the upstroke in systole, with $k_3$, $\ell_{rr}$, and $r_{min}$ becoming more influential in systole. Interestingly, the values of $S_{T,i}$ spike during the flow upstroke without large increases in $S_i$, suggesting that there are a high level of parameter interactions during flow upstroke. 
In the RPA, there is a large increase in $S_i$ for the parameter $k_3$ during the upstroke of systole, with a similar increase in $S_{T_i}$ as well, suggesting the dominance of this parameter in RPA flow upstroke. 
The RPA also shows a delayed sensitivity for $\ell_{rr}$, which peaks later than in the LPA. The RPA is longer than the LPA, hence this different in $\ell_{rr}$ may be due to wave propagation and reflection differences due to vessel length. Overall, the findings from Figures \ref{fig:Sobol_PCA}(b,d,f) and \ref{fig:Sobol_ST_time} suggest that the structured tree parameters are most influential on average, with relatively less sensitivity to $k_3$ than the Windkessel boundary condition results. This suggests that the structured tree parameters should all be considered for inference, possibly with $k_3$, while again $k_1$ and $k_2$ should be considered fixed.

\subsection*{Experimental designs and profile-likelihood}
Sensitivity analysis can provide information about parameter influence but generally cannot determine whether parameters are identifiable. The only exception is when parameters are considered functionally non-influential on an output \cite{Smith2024} in which case they have no effect on the output of interest. Sensitivity based fixing of non-influential parameters is typically problem and user dependent \cite{Alexanderian2020,Colebank2019,Huberts2014} with cutoff values for model sensitivity difficult to determine. Here we leveraged the use of profile-likelihood confidence intervals, which allow for more formal investigations of model sensitivity. While considered the gold standard for practical identifiability \cite{Renardy2021,Wieland2021,Colebank2022}, these methods are computationally expensive and rarely used with PDE models \cite{Renardy2022structural,Ciocanel2024parameter}. We thus leverage the use of spectral surrogates to provide both sensitivity and identifiability metrics for our PDE model, and further inspect how different experimental designs may lead to different parameter identifiability. The designs chosen reflect possible clinical measurements for the management of pulmonary vascular disease, such as pulmonary hypertension \cite{Colebank2022,Paun2020}, and provide insight into how changing the available data may impact the inference of parameters.

Our results in Figure \ref{fig:PL_WK}(a) support the notion that $k_1$ and $k_2$ are functionally non-identifiable for the Windkessel boundary conditions. Even with the most complex design $D3$, these two parameters show minimal effects on the likelihood function and thus make them impractical to infer. After fixing $k_1$ and $k_2$, we still find that the compliance parameter $C_{T,2}$ is not practically identifiable within the prescribed parameter bounds in Figure \ref{fig:PL_WK}(b); however, we anticipate that increasing the upper bound will lead to this parameter having finite confidence bounds. Thus, if larger values of $C_{T,i}$ are appropriate for a given problem, we consider the parameter set excluding $k_1$ and $k_2$ to be identifiable when using pressure and flow data (i.e., $D3$). We also considered fixing one set of Windkessel parameters (in the LPA) and inferring the parameters of the RPA to see if there were inherent parameter dependencies, and Figure \ref{fig:PL_WK}(c) does show that designs $D1$ and $D2$ are more informative when LPA parameters are fixed. This suggests that the LPA and RPA parameters interact with each other and cause issues with identifiability, whereas inferring one set of Windkessel parameters reduces these identifiability issues. This is consistent with findings in prior studies by Colebank et al. \cite{Colebank2019}, Qureshi et al. \cite{Qureshi2019}, and Paun et al. \cite{Paun2020}, which used a reduce set of Windkessel scaling factors instead of the full set of resistance and compliance elements to overcome identifiability issues.

Evaluations of the model along the profile-likelihood, shown in Figure \ref{fig:PL_eval_WK}, show how parameter confidence bounds translate to output signals. The low sensitivity of $k_1$ and confidence bound below the confidence threshold correspond to a non-identifiability, and pressure simulations along the profile likelihood overlay. Area and flow predictions (not shown) also overlay for designs $D2$ and $D3$, respectively, again reflecting the effects of non-identifiable parameters on output signals. Results for $k_3$ and $R_d1$ show how strong identifiability for a given parameter (i.e., narrow confidence bounds) correspond to output signals that deviate from the true data when constructing the profile-likelihood.

The analysis for the structured tree model in Figure \ref{fig:PL_ST} is qualitatively distinct from the Windkessel results in Figure \ref{fig:PL_WK}. In particular, the full parameter set (Figure \ref{fig:PL_ST}(a)) of the structured tree model has numerous local minima and appears riddled with identifiability issues. It has been documented elsewhere \cite{Paun2020,Colebank2024,TaylorLapole2025} that the stiffness and structured tree parameters are interdependent on the simulated hemodynamics, and hence cause issues when calculating the profile-likelihood. To proceed, we use the low sensitivity of $k_1$ and $k_2$ as a starting point for parameter fixing in Figure \ref{fig:PL_ST}(b). The profile-likelihood confidence intervals clearly improve for several parameters, especially when using $D3$, but still show identifiability issues. In particular, $\alpha$ and $\beta$ appear to oscillate between minima, while $r_{min}$ appears non-identifiable even for $D3$. These results collectively lead to Figure \ref{fig:PL_ST}(c) and, after seeing that $\alpha$ and $\beta$ are not jointly identifiable, lead to a final three dimensional parameter space including $k_3$, $\alpha$, and $\ell_{rr}$ in Figure \ref{fig:PL_ST}(d). This is a similar parameter set as those determined by Taylor-Lapole et al. \cite{TaylorLapole2025} and Paun et al. \cite{Paun2025}, which used sensitivity analysis to reduce the structured tree model dimensionality. We again see that $D3$ by far is the most informative design, and provides evidence that flow data to the left and right pulmonary trunk are informative in inferring model parameters. Model evaluations along the profile likelihood in Figure \ref{fig:PL_eval_ST} support this conclusion as well, again illustrating how data constraints can cause parameter combinations that greatly affect the likelihood, output signals, and parameter identifiability.

Given that the results are contingent on the use of spectral surrogates, we evaluated the true PDE simulation along the profile-likelihood to ensure that the likelihood geometry and conclusions regarding identifiability are not biased on the emulator. The results for the Windkessel boundary conditions in Figure \ref{fig:PL_simulator_WK} show that, across the three experimental designs, the emulator predicted likelihood and confidence intervals are consistent with simulator derived values. We note that in $D2$, there is a clear bias in the minimum for $R_{p,2}$, though the shape and confidence bounds for the parameter are similar. The overlap between simulator and PCE derived confidence intervals provide evidence that likelihood evaluations by the PCE are consistent. In the case of the structured tree boundary conditions, we again see that there are multiple minima in the likelihood landscape, some of which are not identified by the PCE emulator. For $D1$ in Figure \ref{fig:PL_simulator_ST}, e.g., we see that the profile-likelihood for $\ell_{rr}$ has a rougher landscape than that predicted by the PCE. However, both surrogate and simulator confidence intervals suggest that $\ell_{rr}$ is identifiable. It is clear that the structured tree boundary conditions are harder to infer from data, and thus require sufficient amounts of data (e.g., flow data in $D3$) to ensure identifiability.

We conclude that, with respect to spectral surrogates, the proposed framework provides insight beyond sensitivity analysis alone for parameter fixing. In particular, employing profile-likelihood based confidence intervals provide a more robust manner in which parameters are prioritized for inference versus fixing. In addition, our results suggest that flow data is substantially informative, and provides the likelihood with structure that favors stronger identifiability among parameters. Thus, if feasible, $D3$ is appropriate for future studies that use clinical data for determining subject-specific parameters from either set of boundary conditions.

\subsection*{Limitations}
There are several limitations to the proposed methods and analysis presented here. We use PCEs as a spectral surrogate for the PDE simulator, which are commonly used in engineering \cite{Alexanderian2020,Nagel2020} and biomedical \cite{Colebank2024,Eck2017,Huberts2014} applications. However, other emulation strategies, such as Gaussian processes \cite{Paun2022,Paun2025,Higdon2008} may provide more accurate surrogate models. Future studies leveraging Gaussian processes or neural networks with the profile-likelihood analysis are warranted. We used dimension reduction for each output quantity to provide a more efficient emulation framework. The PCEs are built on the PCA representation for each quantity of interest, yet there is an inherent correlation structure across outputs as well. This can be overcome by considering multioutput emulation where the covariance structed between outputs is accounted for \cite{Higdon2008,GPemulation2}. We construct profile-likelihood confidence intervals in the absence of measurement noise, instead using the inherent error between the simulator and emulator as our measurement variance. Future applications of this approach with real data will require additional error structure, including a separation between emulator error, simulator discrepancy, and measurement error \cite{Paun2020}. Nevertheless, this proof of concept study provides evidence that spectral surrogates can be leveraged beyond sensitivity analysis for formal parameter identifiability analysis. Finally, we examine three experimental designs for inference in our pulmonary hemodynamics model, which align with possible experimental or clinical measurement protocols \cite{Chambers2020,Qureshi2014,Yang2019}. We consider these designs strictly in terms of parameter identifiability, though there are additional criterion (e.g., optimal experimental design \cite{Alexanderian2021}). These approaches should be used in parallel with parameter identifiability analysis.

\section*{Conclusions}
This work combines spectral surrogates with the profile-likelihood confidence interval approach to determine model sensitivity and parameter identifiability. We provide evidence that PCEs can be used beyond sensitivity analysis, providing an approach that assesses model sensitivity via Sobol' indices and then further exploits the use of PCEs as a emulator for constructing profile-likelihood confidence intervals. We assess vector output sensitivity through dimension reduction and provide both PCA-based and pointwise-in-time global sensitivities for the hemodynamics model. Our results show that parameter sensitivity can be used in combination with formal identifiability analyses for a more robust parameter fixing approach. Moreover, we show that identifiability analysis and changes in parameter identifiability with different experimental designs can determine whether inverse problems will have unique parameter solutions. In total, this work provides a pipeline for emulator based analyses that can be used for experimental planning when using complex PDE simulators in biomedical settings.

\section*{Supporting information}


\paragraph*{S1 File.}
\label{S1_File}
{\bf Profile-likelihood results.}  Additional profile-likelihood results for the five test data sets.

\section*{Data Availability Statement} 
The source code for simulations and analyses can be found at \url{https://github.com/mjcolebank/Colebank_Identifiability_PDE}.


\nolinenumbers

%
%
%





\bibliography{references}

\end{document}